\documentclass[fleqn,usenatbib]{mnras}

\usepackage{newtxtext,newtxmath}

\usepackage[T1]{fontenc}

\DeclareRobustCommand{\VAN}[3]{#2}
\let\VANthebibliography\thebibliography
\def\thebibliography{\DeclareRobustCommand{\VAN}[3]{##3}\VANthebibliography}

\usepackage{graphicx}	
\usepackage{amsmath}	

\usepackage{multirow}
\usepackage{longtable}

\title[Dust-obscured SMFs from $z\sim3-8$ with JWST]{Unveiling the hidden universe with JWST: The contribution of dust-obscured galaxies to the stellar mass function at $\mathbf{z\sim3-8}$}

\author[R. Gottumukkala et al.]{R. Gottumukkala,$^{1,2,3}$\thanks{E-mail: rashmi.gottumukkala@gmail.com}
L. Barrufet,$^{1,4}$
P. A. Oesch,$^{1,2,3}$
A. Weibel,$^{1}$
N. Allen,$^{2,3}$
B. Alcalde Pampliega,$^{5}$
\and
E. J. Nelson,$^{6}$
C. C. Williams,$^{7,8}$
G. Brammer,$^{2,3}$
Y. Fudamoto,$^{9,10}$
V. Gonz\'alez,$^{11}$
K. E. Heintz,$^{2,3}$
\and
G. Illingworth,$^{12}$
D. Magee,$^{12}$
R. P. Naidu,$^{13}$
M. Shuntov,$^{2,3}$
M. Stefanon,$^{14,15}$
S. Toft,$^{2,3}$
\and
F. Valentino,$^{2,3,16}$
M. Xiao$^{1}$
\\
\\
$^{1}$Department of Astronomy, University of Geneva, Chemin Pegasi 51, 1290 Versoix, Switzerland\\
$^{2}$Cosmic Dawn Center (DAWN)\\
$^{3}$Niels Bohr Institute, University of Copenhagen, Jagtvej 128, DK-2200, Copenhagen N, Denmark\\
$^{4}$Institute for Astronomy, School of Physics \& Astronomy, University of Edinburgh, Royal Observatory, Edinburgh, EH9 3HJ, UK\\
$^{5}$European Southern Observatory, Alonso de C\'{o}rdova 3107, Vitacura, Santiago, 
Regi\'{o}n Metropolitana, 7630355, Chile\\
$^{6}$Department for Astrophysical and Planetary Science, University of Colorado, Boulder, CO 80309, USA\\
$^{7}$NSF's National Optical-Infrared Astronomy Research Laboratory, 950 North Cherry Avenue, Tucson, AZ 85719, USA\\
$^{8}$Steward Observatory, University of Arizona, 933 N. Cherry Avenue, Tucson, 85721, USA\\
$^{9}$Waseda Research Institute for Science and Engineering, Faculty of Science and Engineering, Waseda University, 3-4-1 Okubo, Shinjuku, Tokyo 169-8555, Japan\\
$^{10}$National Astronomical Observatory of Japan, 2-21-1, Osawa, Mitaka, Tokyo, Japan\\
$^{11}$Departamento de Astronom\'ia, Universidad de Chile, Camino del Observatorio 1515, Las Condes, Santiago 7591245, Chile\\
$^{12}$Department of Astronomy and Astrophysics, University of California, Santa Cruz, CA 95064, USA\\
$^{13}$MIT Kavli Institute for Astrophysics and Space Research, 77 Massachusetts Ave., Cambridge, MA 02139, USA\\
$^{14}$Departament d'Astronomia i Astrof\`isica, Universitat de Val\`encia, C. Dr. Moliner 50, E-46100 Burjassot, Val\`encia,  Spain\\
$^{15}$Unidad Asociada CSIC ``Grupo de Astrof\'isica Extragal\'actica y Cosmolog\'ia'' (Instituto de F\'isica de Cantabria - Universitat de Val\`encia)\\
$^{16}$European Southern Observatory, Karl Schwarzschild Strasse 2, 85748 Garching, Germany\\
}
\date{Accepted XXX. Received YYY; in original form ZZZ}

\pubyear{2024}

\begin{document}
\label{firstpage}
\pagerange{\pageref{firstpage}--\pageref{lastpage}}
\maketitle

\begin{abstract}
With the advent of JWST, we can probe the rest-frame optical emission of galaxies at $z>3$ with high sensitivity and spatial resolution, making it possible to accurately characterise red, optically-faint galaxies and thus move towards a more complete census of the galaxy population at high redshifts.
To this end, we present a sample of 148 massive, dusty galaxies from the JWST/CEERS survey, colour-selected using solely JWST bands. With deep JWST/NIRCam data from 1.15$\mu$m to 4.44$\mu$m and ancillary HST/ACS and WFC3 data, we determine the physical properties of our sample using spectral energy distribution fitting with BAGPIPES. We demonstrate that our selection method efficiently identifies massive ($\mathrm{\langle \log M_\star/M_\odot \rangle \sim 10}$) and dusty ($\mathrm{\langle A_V\rangle \sim 2.7\ mag}$) sources, with a majority at $z>3$ and predominantly lying on the galaxy main-sequence.
The main results of this work are the stellar mass functions (SMF) of red, optically-faint galaxies from redshifts between $3<z<8$: these galaxies make up a significant relative fraction of the pre-JWST total SMF at $3<z<4$ and $4<z<6$, and dominate the high-mass end of the pre-JWST SMF at $6<z<8$, suggesting that our census of the galaxy population needs amendment at these epochs.
While larger areas need to be surveyed in the future, our results suggest already that the integrated stellar mass density at $\mathrm{\log M_\star/M_\odot\geq9.25}$ may have been underestimated
in pre-JWST studies by up to $\sim$15-20\% at $z\sim3-6$, and up to $\sim$45\% at $z\sim6-8$, indicating the rapid onset of obscured stellar mass assembly in the early universe.
\end{abstract}

\begin{keywords}
galaxies: high-redshift -- galaxies: evolution -- infrared: galaxies -- methods: observational -- techniques: photometric
\end{keywords}



\section{Introduction}
\label{sec:introduction}

For decades, observational astronomers have been on a quest to determine how the galaxy population evolves through cosmic time.
The Hubble Space Telescope (HST) has pioneered the study of this question: HST has observed high-redshift galaxies, primarily through their rest-frame ultraviolet (UV) emission. These high-redshift galaxies, usually referred to as `normal' or Lyman-break galaxies (LBGs), have been studied extensively from $\mathrm{z\sim3}$ to $\mathrm{z\sim11}$, tending to have moderate star formation rates (SFR) and stellar masses, and are thought to make up the bulk of the galaxy population \citep[e.g.,][]{Labbe2013, Schaerer2013,  Bouwens2015, Finkelstein2015, Oesch2016, Faisst2020}. These mostly dust un-obscured galaxies are also thought to dominate the cosmic star-formation rate density (SFRD) at $\mathrm{z>4}$, while at lower redshifts the universe was dominated by obscured star-formation \citep[e.g.,][]{Madau2014,Zavala2021}. While `normal', un-obscured galaxies have been well-studied, our census of the galaxy population remains incomplete at $\mathrm{z > 3}$ as rest-frame UV selections systematically miss massive, dust-obscured sources \citep[e.g.,][]{Pampliega2019,Wang2019}.

Over the last decade, a significant population of optically undetected galaxies with relatively bright infrared (IR) or sub-millimetre (sub-mm) emission has been discovered, several in Spitzer/IRAC data and some of them with ALMA detections \citep[e.g.,][]{Huang2011, Simpson2014, Caputi2015, Stefanon2015, Wang2016, Franco2018, Pampliega2019, Wang2019, Yamaguchi2019, Williams2019, Dudzeviciute2020, Sun2021, Smail2021, Manning2022,  Shu2022, Xiao2023a}.
They typically have very red spectral energy distributions (SEDs) and remain undetected even in deep HST $H$-band observations -- hence their name: HST-dark galaxies. Their SEDs are not well-constrained, with a few photometric detections and lack of spectroscopic redshifts, which result in very large uncertainties on their photometric redshifts, stellar masses, and SFRs \citep[e.g.,][]{Caputi2012, Stefanon2015,  Williams2019, Pampliega2019}. The physical properties of these galaxies were largely unconstrained until the arrival of the James Webb Space Telescope \citep[JWST,][]{Gardner2023}.

JWST has revolutionised the field of optically-faint galaxies, providing for the first time reliable physical parameters \citep[e.g.,][]{Barrufet2023, Nelson2023_pub, Gonzalez2023, Rodighiero2023, Labbe2023a, Guijarro2023_pub}. With its unprecedented sensitivity and resolution in the near-IR, JWST probes the rest-frame optical emission of galaxies at $z\gtrsim3$, allowing one to identify the Balmer break, a good redshift and mass indicator.
Additionally, the SEDs of massive galaxies are typically highly dust-attenuated with characteristic red slopes in the rest-frame optical. With its extensive photometric coverage from $1-5~\mu$m, JWST's Near-Infrared Camera \citep[NIRCam;][]{Rieke2023} is the ideal instrument to identify sources based on these features.

The early JWST era has seen the puzzling emergence of two additional populations of galaxies. The first is a population of massive sources ($\mathrm{>10^{10}\ M_\odot}$) at $z>7$, less than 700 Myr after the Big Bang \citep[e.g.,][]{Labbe2023a}. With the currently accepted theory of hierarchical structure formation within $\Lambda$CDM cosmology, it is challenging to explain how galaxies could accumulate this much mass through mergers or accretion alone \citep{Kolchin2023, Menci2022}, while it might still be possible to reconcile such observations with theory \citep{Mason2023,Dekel2023}. One possibility is that these sources are actually active galactic nuclei (AGN), with one \citet{Labbe2023a} source being spectroscopically confirmed to be an AGN with broad emission lines \citep{Kocevski2023_pub}. A deeper investigation into massive galaxies in the early universe is needed in order to determine their abundance and place constraints on mass assembly.

The second emergent population consists of massive quiescent galaxies at high redshifts, now spectroscopically confirmed up to $z=4.658$ \citep{Carnall2023_pub}.
Relatively little physical insight has been provided by simulations thus far to explain the emergence of quiescent galaxies at $z>3$, with simulations struggling to predict observed number densities \citep{Valentino2023_pub,Gould2023_pub}.
While it is highly likely that sub-millimetre galaxies (SMGs) evolved into massive quiescent galaxies at $z\sim2$ \citep{Toft2014}, their number densities are insufficient to explain the presence of quiescent galaxies at $z\sim3-4$ \citep{Valentino2020, Valentino2023_pub}.
Hence, an important step towards understanding the emergence of quenched galaxies is to look for previously-missed massive, dusty galaxies in the early universe and determine their stellar masses and abundances.

For the study of galaxy abundances, the stellar mass function (SMF) is an extremely useful statistical tool to quantify the evolution of the galaxy population as a function of stellar mass across cosmic history.
Determining the SMF at various epochs in the history of the universe allows us to track early galaxy build-up. Several studies have so far constrained high-$z$ SMFs with ground- and space-based multi-wavelength observations \citep[e.g.,][]{Davidzon2017, Stefanon2015, Stefanon2017b, Stefanon2021, McLeod2021, Santini2021, Weaver2023_smf, NavarroCarrera2023}, with the shape of the total SMF being found to be accurately described by the empirically motivated \cite{Schechter1976} function. Given that JWST is primed to find massive, dust-obscured sources that have previously been missed in the galaxy census, this raises the question of whether or not the total SMF at high-$z$ epochs requires modification. The central question we aim to address with this work is,
`How do massive, dusty galaxies selected with JWST affect the high-mass end of the galaxy stellar mass function in the early universe?'

In this study, we use data from the Cosmic Evolution Early Release Science (CEERS) survey \citep{Finkelstein2022b,Finkelstein2023}, a JWST Cycle 1 community survey in the CANDELS/EGS field. CEERS is aimed at discovering the first galaxies and observing galaxy assembly at $z>3$. Given its deep photometric coverage with JWST/NIRCam from 1.15$\mu$m - 4.44$\mu$m, CEERS is the ideal survey to look for red, IR-bright galaxies.

This paper is structured as follows: In Section \ref{sec:sample_selection}, we discuss the photometric data used from HST and JWST and the production of the HST-JWST merged photometric catalogue. We introduce our colour selection using photometry solely from JWST. Furthermore, we describe how we create an AGN-cleaned sample of purely star-forming galaxies. In Section \ref{sec:SED_fitting}, we explain the SED fitting performed using the {\tt Python} tool BAGPIPES \citep{Carnall2018}. In Section \ref{sec:results1}, we discuss the physical properties of our sample, and situate our galaxies on the galaxy main-sequence (\ref{subsec:galaxy_ms}). In Section \ref{sec:results2_smf}, we discuss the methodology used to compute the SMFs (\ref{subsec:determining_smf}) and present the SMFs of massive, dusty galaxies at $3<z<4$, $4<z<6$, and $6<z<8$ (\ref{subsec:smf_final}). Finally, we discuss our sample in the context of other JWST studies in Section \ref{sec:discussion}, and we summarise and conclude our study in Section \ref{sec:summary_conclusion}.

For this work, we assume a flat $\Lambda$CDM cosmological model with H$_0 = (67.8 \pm 0.9)\ \text{km}\ \text{s}^{-1}\ \text{Mpc}^{-1}$ and $\Omega_m = 0.308 \pm 0.012$ as found by the \cite{Planck2016}. All magnitudes are quoted in the AB magnitude system \citep{Oke1983}.
Throughout this paper, we use a \cite{Kroupa2001} initial mass function (IMF). If required for comparison, we scale mass values used in the literature from \cite{Salpeter1955} or \cite{Chabrier2003} to \cite{Kroupa2001} using the scale factors quoted in \cite{Madau2014}.

\section{Observations and sample selection}
\label{sec:sample_selection}

In this section, we describe the imaging data used in this work and the production of the HST and JWST merged photometric catalogue for the CEERS field. In addition, we present our sample selection criteria in order to identify massive and dusty galaxies, including the colour-selection we develop as well as the criteria used to identify and remove AGN from our final sample.

\subsection{Imaging data}
\label{subsec:imaging_data}

We use data from the Cosmic Evolution Early Release Science (CEERS) programme, one of JWST's first early-release science surveys in Cycle 1, with data collected in June and December 2022 \citep{Finkelstein2022b,Finkelstein2023}. CEERS comprises 10 NIRCam pointings covering $\sim$100 arcmin$^2$ in the Extended Groth Strip (EGS) field, a CANDELS legacy field containing a wealth of ancillary HST multi-wavelength data. The NIRCam data covers a range of wavelengths from 1.15$\mu$m to 4.44$\mu$m in the following filters: F115W, F150W, F200W, F277W, F356W, F410M, and F444W (where W and M indicate a wide or medium band filter). Ancillary HST data from the ACS imager is available at wavelengths between 435nm to 814nm (in 3 filters: F435W, F606W, and F814W) and from the WFC3 imager at wavelengths between 1.05$\mu$m to 1.60$\mu$m (in 4 filters: F105W, F125W, F140W and F160W) \citep{Koekemoer2011, Grogin2011, Stefanon2017a}.

For this work, we use the version 5 images reduced with the \texttt{grizli} pipeline and made publicly available by G. Brammer\footnote{\url{https://dawn-cph.github.io/dja/}}, following the same steps as outlined in \cite{Valentino2023_pub}.
The images include all available data over these fields taken with HST and JWST. The imaging depths as measured in circular apertures with a radius of 0.16\arcsec are listed in Table \ref{tab:5sig_depths}. They vary between 28.6 mag to 29.2 mag in the JWST wide filters and are $\sim28.3$ mag in the shortest wavelength ACS imaging.

\subsection{Production of the HST-JWST photometric catalogue}
\label{subsec:catalogue_production}

We use the JWST and ancillary HST images to create photometric catalogues, taking into account the wavelength-dependent point-spread function (PSF). In the following, we briefly describe how the PSF-matched photometric catalogue used in this work was produced (see Weibel et al. in prep. for details).

We match the fluxes in all HST+JWST filters to the PSF resolution in the reddest JWST/NIRCam filter, F444W. For the NIRCam and WFC3 filters, we use the PSFs provided by G. Brammer for use with the CEERS grizli mosaics \citep{Brammer2018}\footnote{\url{https://github.com/gbrammer/grizli-psf-library}}.

For the ACS filters, we derive effective PSFs from the science images by first identifying bright, but unsaturated stars without bright neighbouring sources or flagged pixels, from a preliminary \texttt{SourceExtractor} run \citep{Bertin1996}. Then, we use the method \texttt{EPSFBuilder} from the python package \texttt{photutils} \citep{Bradley2022} which is based on the model developed by \citet{AndersonKing2000} to obtain the final effective PSFs.

We compute matching kernels from each ACS and NIRCam PSF to the NIRCam/F444W PSF using the software package \texttt{pypher} \citep{Boucaud2016} and convolve each flux and root mean square (rms) image with the respective kernel to match the PSF resolution in F444W.

We follow a different procedure for the WFC3 filters because their PSFs are broader than the NIRCam/F444W PSF. First, we compute matching kernels from all of them and from the F444W PSF to the WFC3/F160W PSF, in the same way as described above, and produce PSF-matched flux and rms images accordingly.

Then, we run \texttt{SourceExtractor} in dual mode, using an inverse-variance weighted stack of the unaltered F277W+F356W+F444W images as the detection image and measuring fluxes in circular apertures with a radius of 0.16\arcsec on the original images, the images that were PSF-matched to F444W as well as the images that were PSF-matched to F160W.
For the final catalogue, we use the flux measurements on the original image in F444W and those on the images PSF-matched to F444W for all other filters, except the WFC3 data. For the latter, we correct the fluxes measured on the original images to match the colour between the respective filter and F444W as measured on the images PSF-matched to F160W.

We scale all fluxes to the flux measured in Kron-like apertures by \texttt{SourceExtractor} in F444W, obtained using the default Kron parameters 2.5 and 3.5. To account for residual flux outside the Kron aperture, we measure the fraction of the energy enclosed by a circular aperture with a radius of $\sqrt{a\,b}\,$\texttt{kron\_radius}, where $a$, $b$ and \texttt{kron\_radius} characterise the Kron-ellipse, on the theoretical F444W PSF obtained from \texttt{webbpsf}, and divide all fluxes by that fraction. Finally, we correct all fluxes for Milky Way foreground extinction using the extinction model from \citet{Fitzpatrick2007} through the python package \texttt{extinction},
using the E(B-V) map outlined in \citet{Schlafly2011}.

To get a more realistic estimate of the rms uncertainty of our flux measurements that accounts for correlated noise, we put down circular apertures with a radius of 0.16\arcsec in 5000 random positions on the ``signal-to-noise" image (i.e., the flux image divided by the rms image). We multiply the uncertainties on all fluxes, measured from the rms map respectively, by the scatter measured among those apertures. This leads to a scaling of the flux uncertainties by $\sim$5 - $\sim$35\% depending on the filter - the largest correction being applied to F115W and the smallest to F444W.

\renewcommand{\arraystretch}{1.3}
\begin{table}
\caption{$5\sigma$ depths measured in 0.16 arcsec apertures in HST/ACS, HST/WFC3, and JWST/NIRCam photometric filters. Depths are quoted in AB magnitudes.}
\centering
\label{tab:5sig_depths}
\begin{tabular}{c|c|c}
Telescope/Instrument & Filter & $5\sigma$ depth [AB mag] \\
\hline
\hline

\multirow{3}{*}{HST/ACS} & F435W & 28.27 \\
 & F606W & 28.36 \\
 & F814W & 28.19 \\ \hline
\multirow{4}{*}{HST/WFC3} & F105W & 27.96 \\
 & F125W & 27.74 \\
 & F140W & 26.99 \\
 & F160W & 27.81 \\ \hline
\multirow{7}{*}{JWST/NIRCam} & F115W & 28.63 \\
 & F150W & 28.65 \\
 & F200W & 28.93 \\
 & F277W & 29.17 \\
 & F356W & 29.17 \\
 & F410M & 28.41 \\
 & F444W & 28.81 \\ 
 
\end{tabular} 
\end{table}

To identify and flag stars we used a flux ratio criterion similar to \citet{Weaver2023_smf}. 
We also flag objects as artefacts that are too small to be real sources (typically left-over bad pixels).
The full CEERS catalogue contains 93,922 sources.  Out of these, we remove 930 sources that are either identified as stars or flagged as artefacts based on the above criteria, resulting finally in 92,992 sources.

\begin{figure}
   \centering
   \includegraphics[width=\hsize]{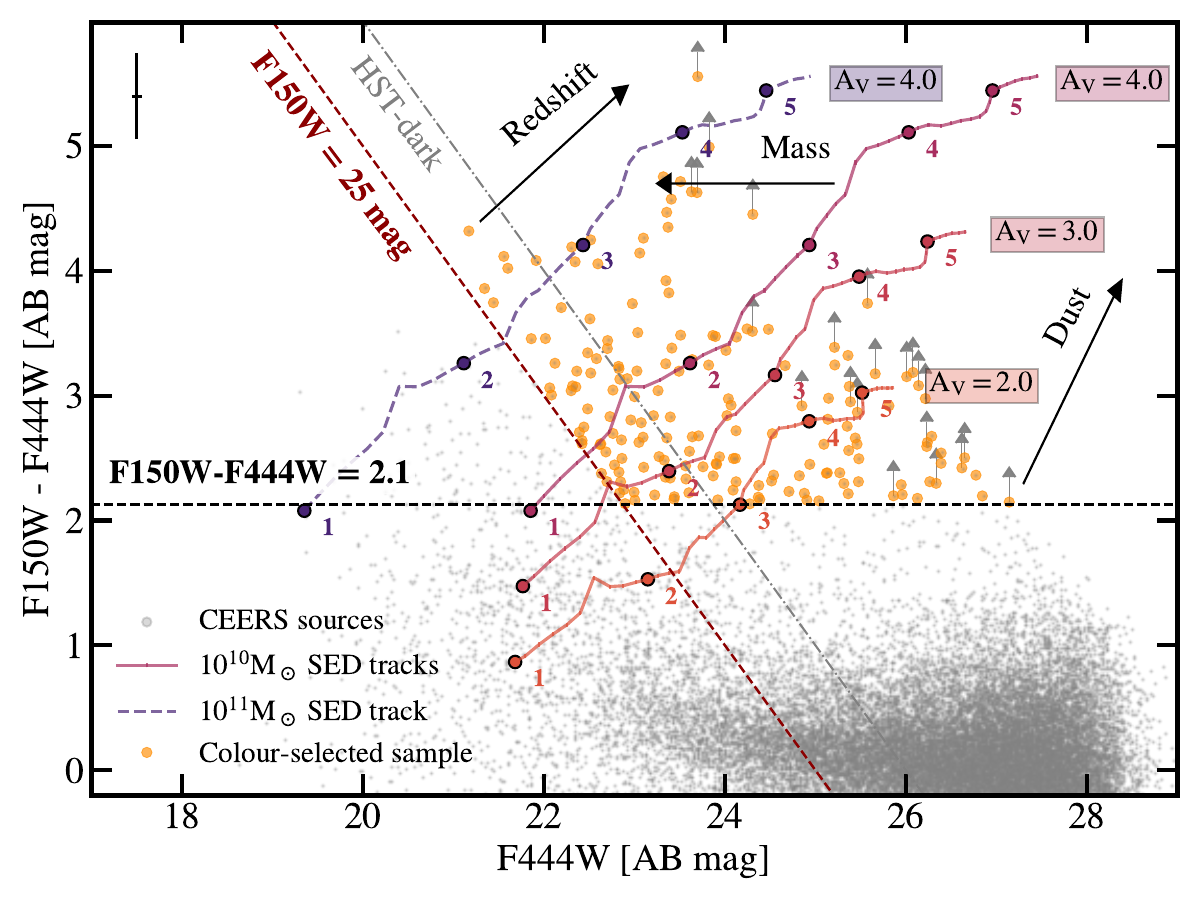}
   \caption{Colour-magnitude diagram of F150W-F444W vs. F444W showing our selection method. The grey scatter points show all the sources in the CEERS catalogue while the orange scatter points are our selected galaxies. The coloured lines are SED tracks for various dust attenuation values ($\mathrm{A_V}$ is indicated in boxes), with coloured numbers indicating the redshift -- solid lines correspond to $\mathrm{10^{10}\ M_\odot}$ and the dashed line corresponds to $\mathrm{10^{11}\ M_\odot}$. The grey arrows show upper limits for the sources with F150W mag lower than 2$\sigma$ (median errors are shown by the cross in the upper right of the figure).
   The colour criterion F150W-F444W$>$2.1 mag (black dashed line) in principle identifies $z>3$ sources with $\mathrm{A_V\gtrsim2\ mag}$ and $\mathrm{\log(M_\star/M_\odot})\sim10$, while the magnitude cut F150W$>$25 mag (dark red dashed line) is designed to rid the sample of low-$z$ sources while retaining the most massive galaxies in the sample ($\mathrm{\sim 10^{11}\ M_\odot}$). Also shown for reference is the F150W = 26 mag cut that is a proxy for identifying HST-dark sources \citep{Gonzalez2023}.
   We select 179 red galaxies with these criteria that theoretically restrict our sample to massive, dusty, high-redshift galaxies.}
   \label{fig:sample_selection}
\end{figure}

\begin{figure*}
   \centering
   \includegraphics[width=\hsize]{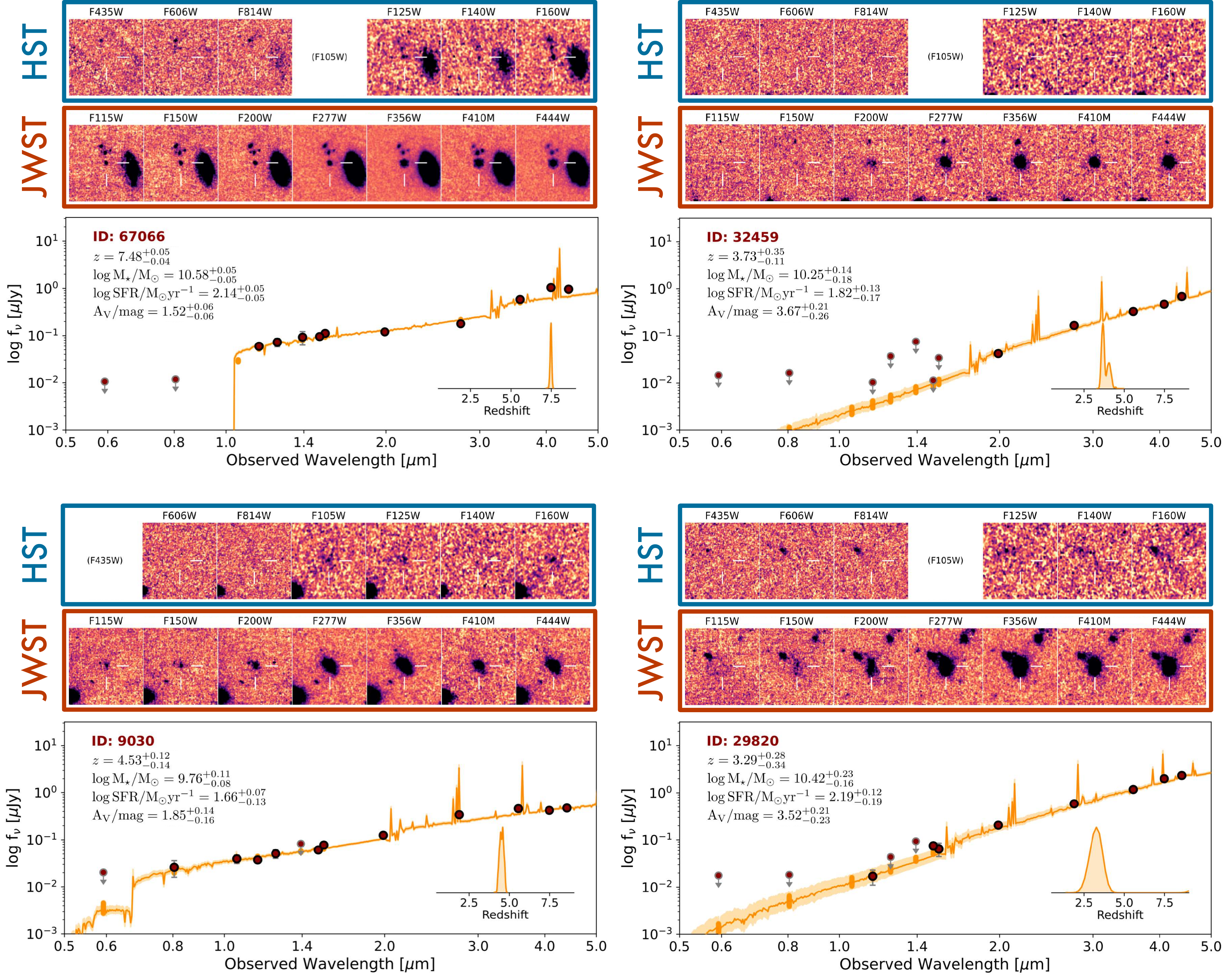}
   \caption{Postage stamps and SED fits of four selected galaxies from our sample of red galaxies. The stamps boxed in blue are from ancillary HST/ACS and WFC3 data, and the stamps boxed in red are from JWST/NIRCam imaging (each stamp is $\mathrm{4\times4\ arcsec^2}$). There is a variety in the morphological properties of our sample, ranging from spatially extended sources to compact ones. The lower panels display the SED fits: the maroon points represent the photometry and the downward arrows represent the flux upper limits.
   The orange lines are the SED fits from BAGPIPES and the photometric redshift probability density functions are inlaid in the lower right part of the graphs. The physical properties of these galaxies are quoted on the graphs. They are massive ($\mathrm{\log M_\star/M_\odot}\gtrsim9.5$) and dusty ($\mathrm{A_V\sim1.5-4\ mag}$) with redshifts ranging from $z\sim3-8$. }
   \label{fig:SEDfit_fnu_wpz}
\end{figure*}

\subsection{Selection of red, optically-dark/faint sources at {\sl z>3}}
\label{subsec:source_selection}

Over the last decade, numerous studies of HST-dark galaxies and red galaxies have been conducted, with dropout and colour selections shown to be effective methods for selecting high-redshift sources. 
Typically, these studies combine HST and Spitzer data to select massive and dusty star-forming galaxies \citep[e.g.,][]{Huang2011,Pampliega2019,Wang2019, Sun2021}.

Several unique colour cuts have been used over the last decade for efficient selections of red galaxies using HST/WFC3 bands in the optical and Spitzer/IRAC and (recently) JWST/NIRCam bands in the near-IR \citep[e.g.,][]{Huang2011,Caputi2012,Wang2016,Pampliega2019,Wang2019,Sun2021,Barrufet2023,Labbe2023a,Nelson2023_pub,Rodighiero2023,Gonzalez2023,Xiao2023a,Long2023a}. Here, we build on these and make a broad selection of red galaxies using solely JWST/NIRCam bands in order to fully exploit the increased sensitivity and resolution of JWST. By designing and implementing a colour selection capable of identifying the effects of the Balmer-break and reddened stellar continuum emission in a galaxy's photometry, we expect to select massive and dusty galaxies at high redshifts.
For this, we use the
{\tt Python} tool Bayesian Analysis of Galaxies for Physical Inference and Parameter EStimation \citep[BAGPIPES,][]{Carnall2018}\footnote{\url{https://bagpipes.readthedocs.io/en/latest/}}
to investigate the evolution of colour with redshift. We generate galaxy spectra, from which we 
extract the photometry and compute modelled colours. We use a delayed-$\tau$ star-formation history, ages of 1 Gyr, an $e$-folding time of 3 Gyr, a mass of $10^{10}$ M$_\odot$ and metallicity of 0.5 Z$_\odot$. We model galaxies at redshifts between $z=(1.,6.)$ in steps of $\Delta z=0.1$ and at discrete dust attenuation values of $\mathrm{A_V=[2.0, 3.0, 4.0]\ mag}$ using a Calzetti dust model \citep{Calzetti2000} to produce the SED tracks of massive, dusty galaxies as shown by the coloured lines in Figure \ref{fig:sample_selection}. We also model a single SED track of a $10^{11}$ M$_\odot$ massive galaxy with $\mathrm{A_V=4\ mag}$.

As the Balmer break gets redshifted beyond 1.5$\mu$m at $z\gtrsim3$, we design a colour cut that requires galaxies to be faint in F150W in comparison to longer wavelength bands. \cite{Gonzalez2023} show with a JWST-selected sample that HST-faint sources extend to higher masses than HST-dark sources. We therefore move beyond the strict HST-dark classification by including HST-faint sources in our selection, so as not to miss the most massive and bright galaxies \citep[HST-dark classification referenced from][]{Gonzalez2023}. 
We also use the F444W band to get the broadest redshift range possible (as the highest redshift sources will have their Balmer break closer to F444W). Given our choice of using the F150W and F444W bands, we identify the F150W - F444W colour at which we expect to select galaxies that are (i) high redshift ($z\gtrsim3$), (ii) massive ($\mathrm{\log M_\star/M_\odot \sim 10}$), and (iii) dusty ($\mathrm{A_V\gtrsim2\ mag}$).
In addition, from the SED-tracks shown in Figure \ref{fig:sample_selection}, we estimate the F150W magnitude at which we rid the sample of low-$z$ sources ($z\lesssim2$) while retaining the most massive and dusty galaxies in our sample.

Using the SED modelling described above, we determine a selection that is optimised to identify galaxies with $\mathrm{A_V\gtrsim2\ mag}$ and $\mathrm{\log M_\star/M_\odot}\sim10$ at $z\gtrsim3$, described in Equation \ref{eq:colour_selection}:

\begin{align} 
\label{eq:colour_selection}
    {\rm F150W - F444W} &> 2.1 \ \text{mag},\\
    {\rm F150W} &> 25 \ \text{mag}.
\end{align}

Additionally, as the prominent feature of our galaxies is their redness, this suggests that they must have significant emission in the long wavelength bands.
To ensure reliable detections, we require $\mathrm{SNR>5}$ in all three wide filters in the long wavelength channels: F277W, F356W, and F444W. Altogether, this colour selection is more flexible than in previous studies \citep[i.e.][]{Barrufet2023}; we later remove the $z<3$ sources after evaluating their physical properties (see Section \ref{subsec:photometric_redshifts}).

We find 179 galaxies that satisfy the ${\rm F150W - F444W > 2.1\ mag}$ and ${\rm F150W > 25\ mag}$ criteria out of the $>$90,000 sources in our catalogue (see Figure \ref{fig:sample_selection}).

\subsection{Identifying and removing obscured AGN}
\label{subsec:agn}

In recent literature, there has been mounting evidence from JWST of a population of high redshift obscured AGN that displays very red colours in the NIR \citep{Labbe2023b,Matthee2023,Barro2023,Greene2023}. These so-called little red dots (LRDs) have characteristically blue rest-UV colours which possibly arise from star-forming regions, and red rest-optical colours that arise from the hot, dusty torus of the AGN \citep{Labbe2023b,Greene2023}. These sources are potential contaminants in selections of red, star-forming galaxies, and it is important to address their presence in our sample.

Based on the colour and compactness criteria outlined in \citet{Labbe2023b} and \citet{Greene2023}, we identify a parent sample of 29 potential AGN candidates. In order to further identify point-like sources, we perform a two-component PSF+S\'{e}rsic fit in the F444W filter using the {\tt GalfitM}\footnote{\url{https://www.nottingham.ac.uk/astronomy/megamorph/}} \citep{Haussler2013,Vika2015} software, identifying sources where the flux associated with the PSF component exceeds the flux associated with the S\'{e}rsic component \citep{Labbe2023b}. We identify 20 sources that satisfy these criteria.

We remove these 20 sources from our sample during analysis (Section \ref{sec:results1} onwards), thus considering a purely star-forming sample of galaxies. Figure \ref{fig:6583_stamp_n_SED} in Appendix \ref{sec:agn_appendix} shows the postage stamps and SED of source 6583, identified as one of the 20 AGN candidates in our sample selection. 
Figure \ref{fig:smf_massive_zbins_full_sample} shows the effect of AGN on the SMF, showing that in particular the SMF at $6<z<8$ is significantly overestimated by including AGN.

\section{SED fitting with BAGPIPES to determine the physical properties of galaxies}
\label{sec:SED_fitting}

To calculate the physical properties of our sample,
we use the {\tt Python} tool BAGPIPES \citep{Carnall2018}. BAGPIPES
is an SED-fitting tool capable of modelling galaxies with various star formation histories (such as delayed-$\tau$, exponential, constant, bursts, etc.) and dust models \citep[][etc.]{Cardelli1989,Calzetti2000,Charlot2000}, using stellar population synthesis (SPS) models \citep{Bruzual2003}. We choose to use a delayed-$\tau$ SFH, which has been shown as an effective SFH to model the bulk of the stellar population, and accurately recover stellar masses \citep{Ciesla2017}. Furthermore, this SFH has been successfully used in previous studies of HST-dark galaxies and massive galaxies \citep{Wang2016,Wang2019,Pampliega2019,Gonzalez2023,Barrufet2023}.

We perform SED-fitting within a broad parameter space, allowing the code to explore the following ranges: redshifts between $z=(0, 10)$, a delayed-$\mathrm{\tau}$ SF history with $\mathrm{\tau=(0.1,9)}$ Gyr, masses in the range $\mathrm{\log M_\star/M_\odot=(6, 13)}$, metallicities between $\mathrm{Z=(0.2, 1.2)\ Z_\odot}$, a Calzetti dust model with $\mathrm{A_V=(0.2,4)\ mag}$, nebular emission with an ionisation parameter of $\mathrm{\log U=-2}$, and a velocity dispersion of 300. The models chosen have been successfully used for similar types of galaxies, being able to fit red SEDs \citep{Wang2016, Wang2019, Barrufet2023}. The broad parameter space in each model allows us to explore this enigmatic galaxy population and unveil their physical properties in more detail, in particular their stellar masses.

To test the suitability of our chosen ${\rm A_V}$ range, we allow ${\rm A_V}$ to vary from (0, 6) mag, finding that some galaxies are fit to very dusty ($\mathrm{A_V>4\ mag}$) solutions at low redshifts ($z<0.75$). Galaxies with similar properties have been reported in \cite{Caputi2012} and more recently in \cite{Bisigello2023_pub}, where BAGPIPES is used. We compare the redshifts from BAGPIPES with redshifts derived from the Easy and Accurate Zphot from Yale (EAZY) software \citep{Brammer2008},
using the \texttt{blue\_sfhz} template set\footnote{\url{https://github.com/gbrammer/eazy-photoz/tree/master/templates/sfhz}}.
We find that the photo-$z$'s of the $\mathrm{A_V>4\ mag}$ sources are not in good agreement with EAZY, where EAZY typically finds higher-$z$ solutions with lower $\mathrm{A_V}$.
This is expected, as the maximum $\mathrm{A_V}$ that EAZY can describe is redshift dependent, reaching a maximum of $\mathrm{A_V}\sim4$ at $z\sim3$.
In addition, upon visual inspection of the postage stamps, we find that several of these sources are very compact, completely dropping out of the shorter wavelength filters and thus being more likely to lie at higher redshifts than at $z<0.75$. The inclusion of MIRI data could potentially  rule out the low-$z$ solutions. However, this is only available over a very small portion of the field currently. We refer to Alcalde Pampliega et al. in prep. for a more detailed analysis including MIRI data.

Additionally, given that our aim is to derive accurate stellar masses in order to calculate the stellar mass function, we test whether the derived stellar masses change significantly if we use the EAZY photometric redshifts as an input to the BAGPIPES SED fitting. We find that with EAZY-$z$ as an input, $\mathrm{\langle \log M_\star/M_\odot \rangle = 10.18_{-0.50}^{+0.40}}$ and with the BAGPIPES-$z$, $\mathrm{\langle \log M_\star/M_\odot \rangle = 10.15_{-0.50}^{+0.43}}$. Both derived stellar masses follow a tight 1:1 relation with an average scatter of 0.2 dex, suggesting that the final stellar mass functions will not be strongly affected by our choice of input redshift. We finally use the BAGPIPES-$z$ in all SED fittings.

Examples of some SED fits are shown in Figure \ref{fig:SEDfit_fnu_wpz}, which showcases the variety in galaxy morphology and physical properties. Most of our sources have very red slopes indicating high dust attenuation. We find a diversity in morphology: some sources are spatially extended, while others are extremely compact (see Figure \ref{fig:SEDfit_fnu_wpz}).

We performed a visual inspection of SEDs and postage stamps for all sources while considering their derived physical properties. We remove 11 sources from our sample due to either clearly overestimated photometric redshifts and masses (spatially extended sources that are likely at lower redshift) or sources with deblending issues.
Our final sample thus contains 148 galaxies.

To recapitulate, out of the colour-selected sample of 179 galaxies outlined in Section \ref{subsec:source_selection}, we remove 20 AGN candidates (described in Section \ref{subsec:agn}) and further remove 11 sources that have poor SED fits, resulting in a final sample of 148 galaxies.

\section{Physical properties of red, optically-faint galaxies}
\label{sec:results1}

JWST's outstanding sensitivity and resolution in the near-IR allow us to determine photometric redshifts and physical parameters (such as stellar masses, star formation rates, etc.) with unprecedented accuracy. 
This will allow us to place tighter constraints on the stellar mass build-up in the early universe.  
In this section, we present the photometric redshifts and physical characteristics of our galaxies as determined with BAGPIPES
(see Table \ref{tab:physical_properties} for a list of the derived physical parameters of our full sample; AGN candidates are denoted as such but removed from the following analysis).

\begin{figure}
   \centering
   \includegraphics[width=0.7\hsize]{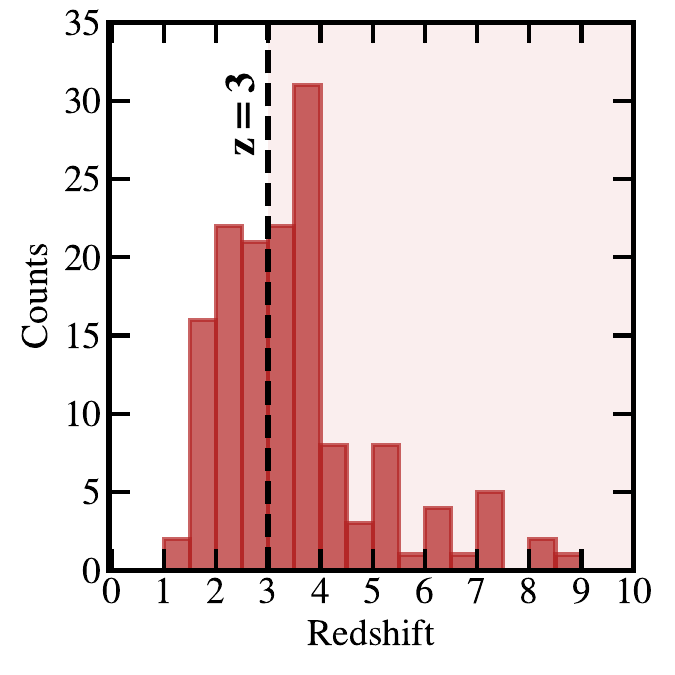}
   \caption{Photometric redshift distribution for 148 red galaxies, determined with the SED fitting tool BAGPIPES. The average redshift is $z_{\rm mean} = 3.46$, with the $\mathrm{16^{th}}$, $\mathrm{50^{th}}$ and $\mathrm{84^{th}}$ percentiles being 2.11, 3.13 and 4.65. $\sim$60\% of the sample lies at $z>3$, reaching $z\sim8$.}
   \label{fig:redshift_hist}
\end{figure}

\subsection{Photometric redshifts}
\label{subsec:photometric_redshifts}

We determine photometric redshifts for our sample of red galaxies using BAGPIPES (see Section \ref{sec:SED_fitting}). The redshift distribution is shown in Figure \ref{fig:redshift_hist}. $\sim$60\% of the sample lies at $z\gtrsim3$ and $\mathrm{\sim 90 \%}$ at $z\gtrsim~2$, with an average redshift of $z_{\rm mean} = 3.46$. This shows that our colour selection successfully identifies high redshifts galaxies, out to $z\sim 8$. The redshift is mostly in agreement with EAZY redshifts using standard templates.

We note the significant number of galaxies that lie at $z<3$ in our selection. We draw the reader’s attention back to Figure \ref{fig:sample_selection}, where we show with the use of arrows that masses and redshifts increase in opposing directions in the colour space of our selection. Further, at a given redshift and stellar mass, there is a scatter in stellar ages and dust which means that invariably, there is a scatter in the properties of the selected population. Therefore, in order to build the most inclusive sample and so as not to miss the most massive and dusty galaxies, it is unavoidable for low-redshift galaxies to enter our selection.

Further, we draw the reader's attention to a caveat of this selection technique, namely the two local peaks seen in the redshift distribution at $z\sim5.5$ and $z\sim7.5$ in Figure \ref{fig:redshift_hist}. The F444W detection is likely driven by the H$\alpha$+[NII] lines at $z\sim4.9-6.6$, and the [OIII]+H$\beta$ lines at $z=6.9-9.0$ \citep[see][]{Oesch2023_pub}. The samples at these redshifts are thus qualitatively different from the bulk sample because their `redness' comes from emission lines rather than the continuum. However, we note that our selection includes $5\sigma$ detection masks in the long-wavelength filters (F277W, F356W and F444W), thus ensuring that the continuum is relatively bright over an extended wavelength range and not just in F444W.
Additionally, as elaborated in the following section, all sources in this study have high $\mathrm{A_V}$ magnitudes; therefore, even if the F444W fluxes of a few select sources are slightly boosted by emission lines, they still qualify as targets for our study. Furthermore, the red and optically faint selection criteria imply that such sources were missing from previous estimates, further justifying their inclusion in our sample.

\subsection{Physical properties of red galaxies}
\label{subsec:physical_properties}

One of JWST's most important improvements in the NIR is its increased photometric coverage at 1-5 $\mu$m in comparison with its predecessor, Spitzer. This allows JWST to better probe the Balmer break and thus derive more accurate photometric redshifts than previously possible. With more accurate photometric redshifts, through SED-fitting we can additionally derive more reliable estimates of the stellar masses of galaxies and their star-formation rates.

\begin{figure}
   \centering
   \includegraphics[width=\hsize]{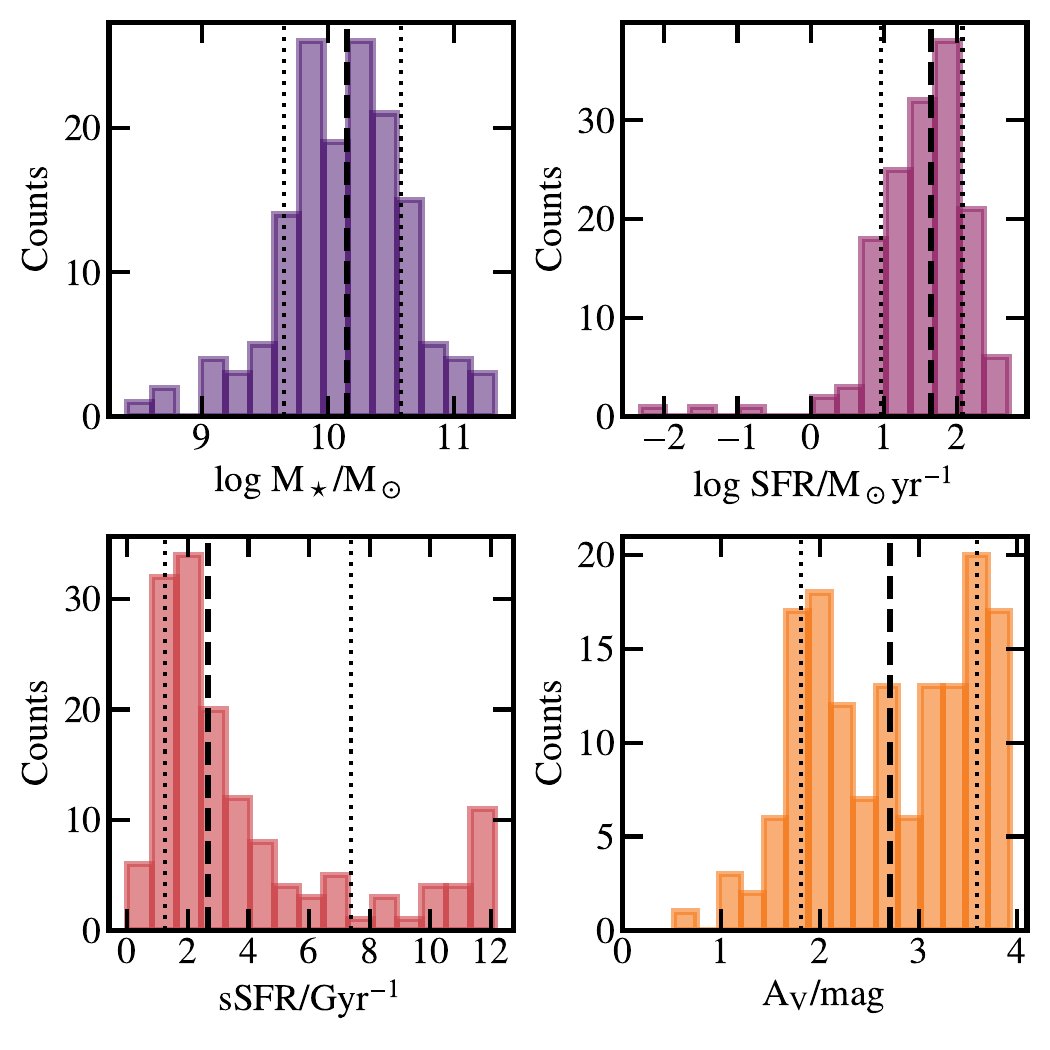}
   \caption{[Top left to bottom right] Histograms of stellar masses, star formation rates, specific star formation rates, and dust attenuations of our sample of 148 red galaxies. The dashed line indicates the $\mathrm{50^{th}}$ percentile while the dotted lines indicate the $\mathrm{16^{th}}$ and $\mathrm{84^{th}}$ percentiles. Our sample is massive ($\mathrm{\langle \log M_\star/M_\odot \rangle = 10.15_{-0.50}^{+0.43}}$) and dusty ($\mathrm{\langle A_V \rangle = 2.71_{-0.91}^{+0.88}\ mag}$), with moderate SFRs of $\mathrm{\langle \log SFR/M_\odot yr^{-1} \rangle = 1.64_{-0.68}^{+0.43}}$, on average below 50 $\mathrm{[M_\odot yr^{-1}]}$ and specific SFRs of $\mathrm{\langle  sSFR/Gyr^{-1} \rangle = 2.66_{-1.42}^{+4.72}}$.}
   \label{fig:red_gal_properties_BAGPIPES}
\end{figure}

We present the distributions of the physical properties of our sample of 148 red, optically-faint galaxies in Figure \ref{fig:red_gal_properties_BAGPIPES}. We find these galaxies to be massive, with a median stellar mass of $\mathrm{\langle \log M_\star/M_\odot \rangle = 10.15_{-0.50}^{+0.43}}$. They also have high dust attenuations of $\mathrm{\langle A_V \rangle = 2.71_{-0.91}^{+0.88}\ mag}$. Additionally, they have moderate star-formation rates, with $\mathrm{\langle \log SFR/M_\odot yr^{-1} \rangle = 1.64_{-0.68}^{+0.43}}$ and $\mathrm{\langle  sSFR/Gyr^{-1} \rangle = 2.66_{-1.42}^{+4.72}}$. As expected, we find our sample to be dominated by relatively massive and dusty star-forming systems.

We note that the SFRs derived in our study are based on rest-frame UV to optical SED fits. We are therefore not modelling the starlight that is reprocessed by dust and emitted in the FIR. While for a more complete picture of the SFR, more FIR data is needed to recover the full infrared SED \citep[see e.g.][]{Xiao2023b}, we refer the reader to \citet{Williams2023}, where they show with a selection of optically-dark galaxies that only those with the most extreme SFRs are significantly affected by the inclusion of MIR and FIR data.

\begin{figure}
   \centering
   \includegraphics[width=\hsize]{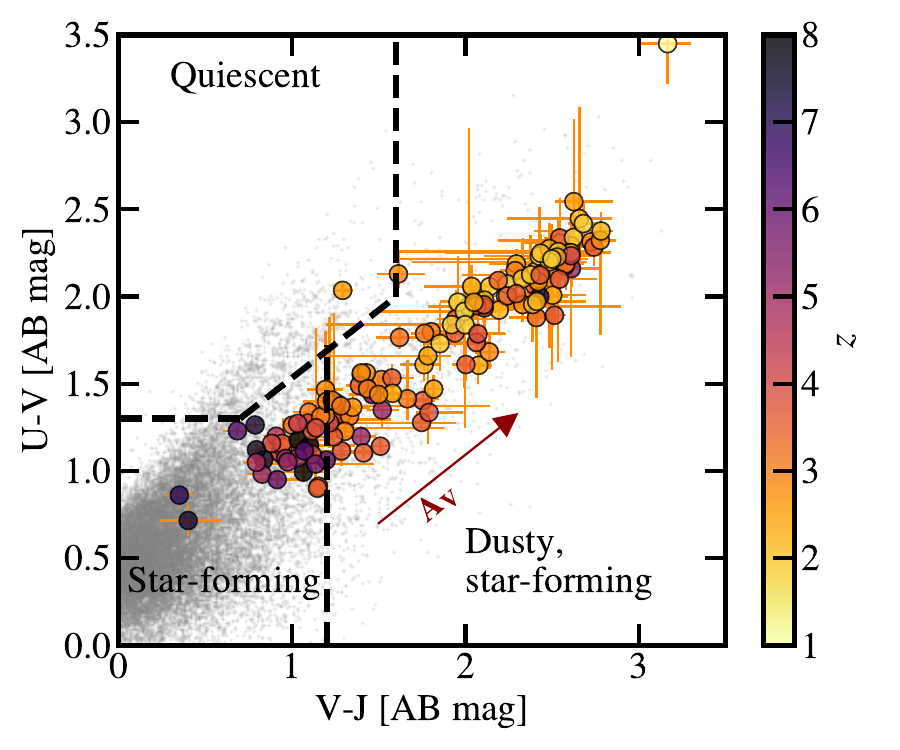}
   \caption{U-V vs. V-J colours of our sample (coloured by redshift), and the CEERS sample (grey scatter points). Uncertainties are given by the $\mathrm{16^{th}}$ and $\mathrm{84^{th}}$ percentiles of the posterior distribution from SED-fitting with BAGPIPES. The galaxy classifications indicated by the black dashed lines are adopted from \citet{Williams2009} and \citet{Spitler2014}.
   The red arrow is the reddening vector, indicating the direction in which the dust attenuation increases.
   All our galaxies (except one) lie in the star-forming regions of the diagram, with $\sim75$\% of the sample lying in the dusty star-forming region. There is a clear tendency for less dusty sources to be at higher redshift.}
   \label{fig:UVJ}
\end{figure}

To further illustrate the dusty nature of our galaxies we situate them on the widely used UVJ diagram.
We classify the star-forming vs. quiescent regions on the UVJ diagram following \cite{Williams2009}, and further split the star-forming region into dusty and unobscured zones following the classification in \cite{Spitler2014}. Figure \ref{fig:UVJ} shows the UVJ classification of our galaxies and of the full CEERS sample. Rest-frame colours for our red galaxies are determined by the best-fit SEDs from BAGPIPES, while for the full CEERS sample they are determined with EAZY due to less expensive computational time. Except for one galaxy lying in the quiescent region of the diagram, the sample lies in the star-forming region, with $\sim 75\%$ of the sample lying particularly in the dusty region. Thus the UVJ classification further indicates the dust-obscured nature of our sample.

\begin{figure}
   \centering
   \includegraphics[width=\hsize]{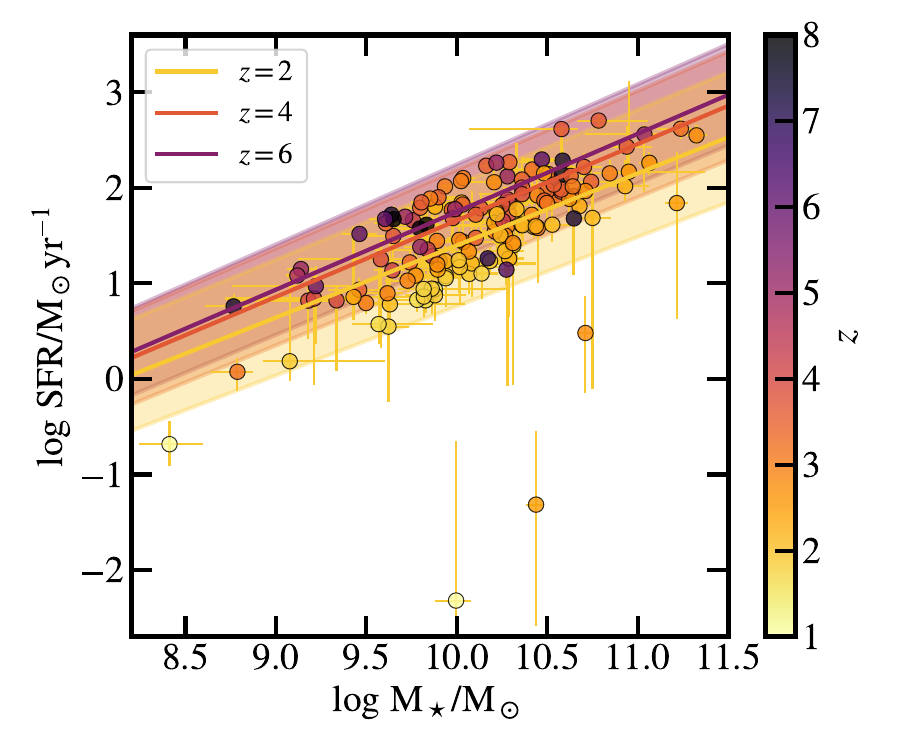}
   \caption{Star formation rate vs. stellar mass for our sample of red, optically-faint galaxies, coloured by photo-$z$ (circular scatter points). Uncertainties are given by the $\mathrm{16^{th}}$ and $\mathrm{84^{th}}$ percentile of the posterior distribution from SED-fitting with BAGPIPES. The galaxy main-sequence lines shown at $z=\mathrm {2,\ 4,\ and\ 6}$ are from \citet{Speagle2014} (solid coloured lines with scatter). The majority of our sample lies on the MS at redshifts of $z<6$, suggesting that they are normal star-forming galaxies with moderate SFRs. The three sources lying significantly below the MS are candidate quiescent galaxies.}
   \label{fig:red_gal_sf_ms}
\end{figure}

\begin{figure}
   \centering
   \includegraphics[width=0.82\hsize]{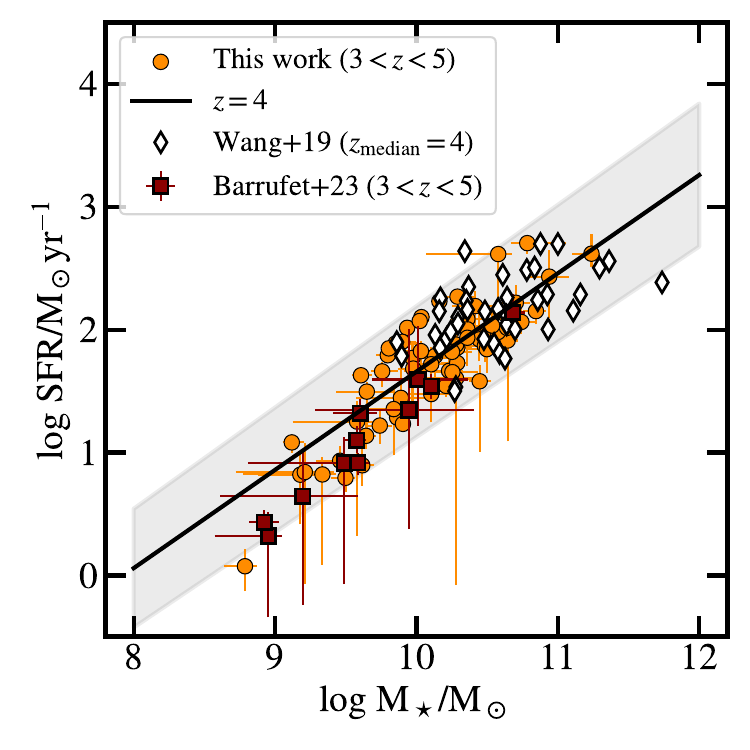}
   \caption{Star formation rate vs. stellar mass for the subset of our sample of red, optically-faint galaxies at $3<z<5$ (circular scatter points). We show the galaxy main-sequence at $z=4.$ from \citet{Speagle2014} (solid line with scatter). We compare our sample to HST-dark galaxies from \citet{Wang2019} with $z_\mathrm{median}=4$ (empty diamonds), and a sample subset from \citet{Barrufet2023} at $3<z<5$ (red squares). Our sample overlaps with the \citet{Barrufet2023} sample at the lower-mass end and with the \citet{Wang2019} sample at the high-mass end, showing that our study covers the mass-range spanned by HST-dark/faint galaxies in both the pre-JWST and JWST era.}
   \label{fig:red_gal_sf_ms_compare_to_lit}
\end{figure}

\subsection{Red galaxies on the galaxy main-sequence}
\label{subsec:galaxy_ms}

To place our galaxies within the context of galaxy evolution, we explore their position on the galaxy main-sequence. Figure \ref{fig:red_gal_sf_ms} shows a plot of SFR vs. $\mathrm{M_\star}$ for our sample, comparing them to the star-forming main-sequence (MS) of galaxies at $z\mathrm{=2,\ 4,\ and\ 6}$ \citep[from][]{Speagle2014}. As shown, our galaxies lie on the star-forming MS, indicative of the `normal' nature of their ongoing star-formation. 
The three galaxies lying significantly below the SF main-sequence are candidate quiescent galaxies at $z<3$. They form less than 2\% of our sample.

We compare our galaxy sample at $3<z<5$ to two studies of interest from the literature: \cite{Wang2019}, that studied ALMA-detected HST-dark galaxies using HST and Spitzer, and the more recent \cite{Barrufet2023}, that studied HST-dark galaxies with HST and JWST. The comparison between samples is shown in Figure \ref{fig:red_gal_sf_ms_compare_to_lit}.

The high-mass end of our sample overlaps with the \cite{Wang2019} sample as our colour selection is inclusive of the \cite{Wang2019} selection criteria. Additionally, the Spitzer/IRAC sensitivity is considerably lower than that of JWST/NIRCam in the same range, thus resulting in the detection of only the brightest and most massive galaxies. We also select lower-mass galaxies than \cite{Wang2019} as JWST can detect galaxies that are fainter in F444W, and our colour selection is less extreme than that used in \cite{Wang2019}.

The low-mass end of our sample overlaps with the range covered by HST-dark galaxies from \cite{Barrufet2023}. This study specifically looked at HST-dark galaxies (F160W $>27$ mag), with JWST/NIRCam's sensitivity permitting detections of lower-mass systems. However, this magnitude cut also limits the detection of brighter, higher-mass sources. By using a less restrictive magnitude cut at 1.5 $\mathrm{\mu m}$ ($\mathrm{F150W>25 mag}$) our selection criteria ensure we find higher-mass galaxies than in \cite{Barrufet2023} while still including the lower-mass HST-dark galaxies in their study. In Figure \ref{fig:red_gal_sf_ms_compare_to_lit}, we show that our sample of red, optically-faint sources lie on the galaxy main-sequence, similar to HST-dark galaxies \citep{Barrufet2023}. The comparison with these select studies from the literature shows that the mass-range spanned by our sample overlaps with both pre-JWST and JWST-selected HST-dark/faint galaxies.

\section{Stellar mass functions of red galaxies: finding the missing sources that dominate the high-mass end}
\label{sec:results2_smf}

In this section, we present the SMFs of red, optically-faint galaxies at redshifts of $3<z<8$. We describe the method used to derive the SMFs and their uncertainties. The SMFs are then presented, discussed, and compared to studies in the literature.

We note that the sample statistics quoted in the previous section were for the full sample of 148 red galaxies across the whole redshift range shown in Figure \ref{fig:redshift_hist}. For the 86 galaxies in the redshift range $3<z<8$, the average stellar masses and dust attenuation values are $\mathrm{ \langle \log M_\star/M_\odot \rangle = 10.17_{-0.56}^{+0.41}}$ and $\mathrm{ \langle A_V \rangle = 2.30_{-0.56}^{+1.22}\ mag}$, setting the stage for the exploration of the SMFs of massive, dust-obscured galaxies at these epochs.

\subsection{Determining SMFs}
\label{subsec:determining_smf}

We use the step-wise method to calculate the SMFs of our sample \citep{Bouwens2008,Santini2021}. The SMFs are approximated by binning the mass distribution, calculating the number of galaxies within each mass bin and dividing this number by the differential comoving volume of the survey. The mass resolution is judiciously chosen to have reasonable statistics within individual mass bins and to have an appropriate mass resolution in order to determine the shape of the SMF.

Given that we detect sources on a stacked image of F277W+F356W+F444W and additionally select sources based on their F150W-F444W colour, we determine the area overlapped by all four filters in the CEERS survey, which is 83.3 $\mathrm{arcmin^2}$. We accordingly calculate the differential comoving volume within the considered redshift bins respectively.

The final SMFs are calculated as shown in Equation \ref{eq:smf}, where $\Phi_{i,j}$ is the estimated number density in a redshift bin `$i$' and mass bin `$j$' per fixed mass bin $\Delta\log {\rm M}$. $N_j$ is the number of galaxies in the $j$'th mass bin, $dV_\text{i, comoving}$ is the differential comoving volume determined within the $i$'th redshift bin and $f_\star$ is a multiplicative factor derived from a completeness simulation used to account for missing sources in our detection catalogues
(described in \ref{subsubsec:smf_completeness}).
\begin{equation}
\label{eq:smf}
    \Phi_{i,j} = N_j\ /\ ( \mathrm{d}V_{i, \text{comoving}} \cdot \Delta\log\mathrm{M}  \cdot f_\star).
\end{equation}

\subsubsection{Completeness}
\label{subsubsec:smf_completeness}

We measure the source detection completeness by running a simple simulation using our custom version of the publicly available software \texttt{GLACiAR2} \citep{Carrasco2018,Leethochawalit2022}. We first select a representative 1.5\arcmin$\times$1.5\arcmin cutout approximately in the middle of the CEERS image with average depth and no contamination by bright stars. Using \texttt{GLACiAR2}, we inject artificial sources, spanning a range of input UV magnitudes from -24.4 to -16.2 in 35 bins at a fixed redshift of $z=6$ into the cutout. We inject 500 sources per bin in batches of maximally 100 sources at a time to avoid overcrowding and run \texttt{SourceExtractor} with the same settings as outlined in Section \ref{subsec:catalogue_production}. The injected galaxies follow a Gaussian distribution in the logarithm of the effective radius, centred at ${\rm 0.8\ kpc}$ and with a scatter of 0.17 dex and they have S\'ersic light profiles with 50\% of the galaxies having a S\'ersic index of 1.5, and 25\% having indices of 1 and 2 respectively. We further assume a flat SED (i.e., a fixed UV-slope of $\beta=-2$), since we only wish to estimate the completeness as a function of apparent magnitude. We repeat this experiment 10 times, therefore injecting 175,000 sources in total.

To obtain the completeness of our sample, we first measure the fraction of recovered galaxies as a function of the input magnitude. Then, for each bin in apparent output magnitude, we determine the completeness as the weighted mean of the completeness values found in each input magnitude bin, weighted by the number of sources from that bin that were observed in the given output magnitude bin. Then, we additionally determine the fraction of detected sources in each apparent magnitude bin that have a measured SNR $>5$ in all of F277W, F356W and F444W (cf. Section \ref{subsec:source_selection}) and multiply that fraction with the detection completeness obtained in the previous step. Since all the observed galaxies considered in this paper have AB-magnitudes $\lesssim27$ in F444W, they are in a regime where the completeness is high and approximately constant as a function of the apparent magnitude (e.g., in F444W). From our analysis, we derive a mean completeness factor of $f_\star=0.87$ by which we scale all our mass functions (see Equation \ref{eq:smf}).

To determine the mass limit above which we are 80\% mass complete, we project our mass distribution onto the SNR limit of our selection \citep[see, e.g.,][]{Pozzetti10}. Given that we select sources that are detected with a 5$\sigma$ certainty in F444W, F356W and F277W, the SNR limit of our selection is $\mathrm{SNR_{lim} = 5\sqrt{3}}$. We calculate the joint SNR for all sources in our sample as $\mathrm{SNR_{joint}^2 = SNR_{F277W}^2 + SNR_{F356W}^2 + SNR_{F444W}^2}$. Assuming that stellar mass values linearly scale with source brightness, we find the hypothetical mass that each source would have if detected at $\mathrm{SNR_{lim}}$: $\mathrm{\log M_{hypothetical} = \log  M_\star - \log ( SNR_{joint} / SNR_{lim})}$. The $\mathrm{80^{th}}$ percentile of the $\mathrm{M_{hypothetical}}$ distribution provides the limit above which the sample is 80\% mass complete, given the specific mass-to-light ratios and SEDs in our sample. We determine that the 80\% mass complete limits are $\mathrm{\log M_\star/M_\odot = 9.15}$ at $3<z<4$, $\mathrm{\log M_\star/M_\odot = 9.07}$ at $4<z<6$ and $\mathrm{\log M_\star/M_\odot = 9.21}$ at $6<z<8$. Therefore, in general, we find that our sample is 80\% mass complete above $\mathrm{M_\star/M_\odot \sim9.25}$ in all redshift bins, and therefore we plot SMFs above this conservative limit. We lose a negligible number of sources by limiting the sample in this manner (1 source each in the redshift bins $3<z<4$ and $6<z<8$).

To consider the completeness of our sample given the flux density limits of the telescope survey, we consider the widely used $V/V_\text{max}$ correction, used to test uniformity in the spatial distribution of sources (\cite{Schmidt1968}, see also \cite{Weaver2023_smf} for a detailed explanation) which particularly affects faint sources. This method considers the maximum redshift, $z_\text{max}$, at which a source within a bin $z_\text{low}<z<z_\text{high}$ would still be observable before falling below the detection limit. Each source is then associated with a maximum observable differential comoving volume, $V_\text{max}$, associated with $z_\text{max}$, and the actual differential comoving volume it is detected in, $V$, associated with $z_\text{high}$. If $z_\text{max}<z_\text{high}$, the source is given a weight of $V/V_\text{max}$, and if $z_\text{max}>z_\text{high}$, $V/V_\text{max}=1$ (as the source would anyways have been detected in the survey, and therefore does not need to be given a higher weightage). Like the step-wise method used to calculate the SMF, the $V/V_\text{max}$ too is non-parametric. It assumes no functional form for the SMF, but it does assume a uniform spatial distribution of galaxies. However, \cite{Weaver2023_smf} show that this is problematic only at $z<1$, thus not affecting our study. We apply the $V/V_\text{max}$ correction to our sources, finding that given the redshift bins we choose, no galaxies in our sample require this correction. This is expected, as our galaxies are red by definition and on average massive and therefore bright in F444W. The $V/V_\text{max}$ correction mostly affects only faint galaxies with the propensity to be detected close to the noise threshold.

\subsubsection{Sources of uncertainty}
\label{subsubsec:smf_unc}

We estimate the uncertainty of the SMFs by considering the Poisson noise $\mathrm{\sigma_N}$, the uncertainty due to cosmic variance $\mathrm{\sigma_{cv}}$, and the systematic uncertainty $\mathrm{\sigma_{sys}}$ due to SED-fitting.

Given that the calculation of the SMF is fundamentally a discrete counting process, the distribution of galaxies within a particular redshift and mass bin must follow Poissonian statistics.
We calculate the uncertainty ${\rm \sigma_N}$ by using frequentist central confidence intervals\footnote{\url{https://docs.astropy.org/en/stable/api/astropy.stats.poisson_conf_interval.html}} \citep[for details, see][]{Maxwell2011}.

An added factor of uncertainty arises from cosmic variance, the field-to-field variation in galaxy number counts due to large-scale structure. It becomes an important source of uncertainty in narrow and deep surveys \citep{Somerville2004}, and is routinely included in uncertainty estimates of the stellar mass function \citep{Davidzon2017,McLeod2021,Weaver2023_smf}. To estimate the cosmic variance $\mathrm{\sigma_{cv}}$, we use the CosmicVarianceCalculator v1.03\footnote{\url{https://www.ph.unimelb.edu.au/~mtrenti/cvc/CosmicVariance.html}} \citep{Trenti2008}, evaluated at the respective number density of our sample.
We find relative cosmic variances for our sample to lie between 20\%-30\%, with the cosmic variance increasing with stellar mass.

\begin{figure*}
   \centering
   \includegraphics[width=\textwidth]{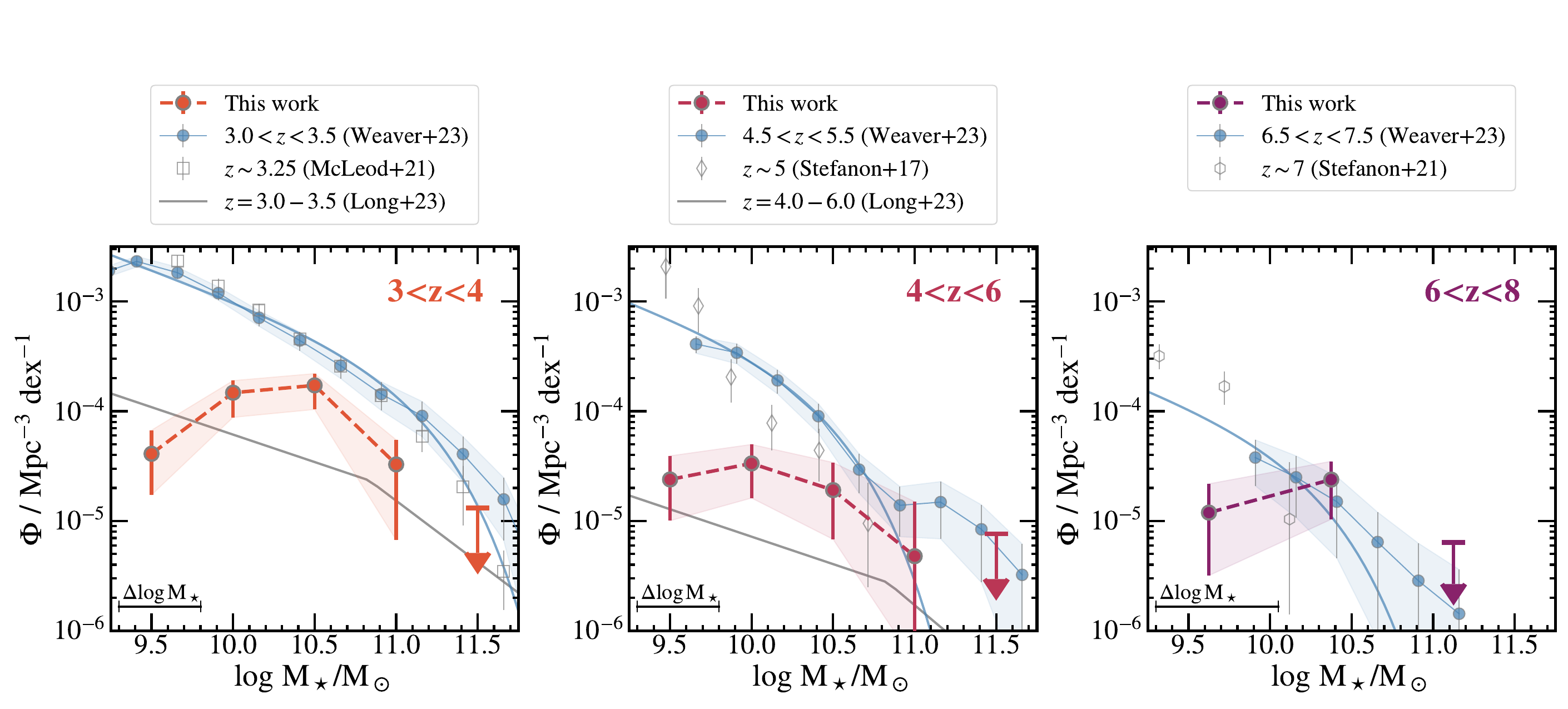}
   \caption{SMFs of our sample of massive and dusty galaxies in three redshift ranges: $3<z<4$, $4<z<6$, and $6<z<8$. Uncertainties shown are derived from Poisson statistics, cosmic variance, and scatter due to SED fitting. Upper bounds (downward arrows) are derived from the right confidence interval of the Poisson distribution. The fixed size of the mass bins are shown in the lower left of each panel. We compare our results to the observed SMFs of the pre-JWST total galaxy population from \citet{Weaver2023_smf} (blue scatter points and shaded area) derived from COSMOS2020 observations \citep{Weaver2023_cosmos} and to the model SMFs from \citet{Long2023} (solid grey line) derived from semi-empirical simulations for dusty star-forming galaxies. For reference, the Schechter fits from \citet{Weaver2023_smf} are also shown (solid blue line). The SMF at $3<z<4$ is additionally compared to \citet{McLeod2021} (square scatter points), derived from ground-based observations, and the $4<z<6$ and $6<z<8$ SMFs are compared to \citet{Stefanon2017b} and \citet{Stefanon2021} at $z=5$ and $z=7$ respectively (diamond and hexagon scatter points), derived from HST and Spitzer imaging.
   At $3<z<4$, comparing our SMF to the pre-JWST \citet{Weaver2023_smf} SMF suggests that up to  $\sim30\%$ of the galaxy population could have been missed at $\mathrm{\log M_\star/M_\odot = 10.5}$ and up to $\sim20\%$ at $\mathrm{\log M_\star/M_\odot = 11.0}$; similarly at $4<z<6$, we find missed fractions of up to $\sim25\%$ at $\mathrm{\log M_\star/M_\odot = 10.5}$ and 11.0. At $6<z<8$, the obscured SMF exceeds the pre-JWST SMF from \citet{Weaver2023_smf} at $\mathrm{\log M_\star/M_\odot = 10.375}$. At both $3<z<4$ and $4<z<6$, our SMFs dominate the dusty model SMF predicted by \citet{Long2023} at $\mathrm{\log M_\star/M_\odot} > 9.5$.}
   \label{fig:smf_massive_zbins}
\end{figure*}

Uncertainties on redshifts and stellar masses can give rise to a scatter, $\mathrm{\sigma_{fit}}$, due to SED fitting. In order to estimate $\mathrm{\sigma_{fit}}$, we generate 1000 independent realisations of the SMF by sampling from the posterior distributions of physical properties derived with BAGPIPES and calculate the variance of the number densities from these realisations. This method provides an estimate of the SMF as well as the uncertainty $\mathrm{\sigma_{fit}}$ on the SMF.

The final uncertainty $\mathrm{\sigma_{tot}}$ of the SMF is the quadrature addition of the Poisson uncertainty, cosmic variance, and the 
uncertainty due to SED fitting \citep[as done in][]{Davidzon2017},
calculated via Equation \ref{eq:smf_unc}:
\begin{equation}
\label{eq:smf_unc}
\mathrm{\sigma_{tot}^2 = \sigma_N^2 + \sigma_{cv}^2 + \sigma_{fit}^2.}
\end{equation}
In the absence of detections, upper limits are calculated as the right confidence interval of the Poisson distribution. This is $1.841 / {( \mathrm{d}V_{i, \text{comoving}} \cdot \Delta\log\mathrm{M} )}$ following \citet{Gehrels1986}.

\subsection{SMFs at $\mathbf{ 3<z<8}$}
\label{subsec:smf_final}

Figure \ref{fig:smf_massive_zbins} and Tables \ref{tab:smf_data_1} and \ref{tab:smf_data_2} present the SMFs of our sample in three redshift ranges: $3<z<4$, $4<z<6$, and $6<z<8$, calculated using the method outlined in the previous section.
We compare our dust obscured SMFs to the observed pre-JWST total SMFs from \citet{Weaver2023_smf}, \citet{McLeod2021} and \citet{Stefanon2021}, derived from ground- and space-based observations. We also compare our SMFs to model dust-obscured SMFs from \citet{Long2023}, which are derived from semi-empirical simulations of dusty star-forming galaxies (DSFGs).

In order to determine the previously {\sl missed} fraction of the SMF, we assume that together, the selection functions of our study and pre-JWST studies produce a more complete survey than solely pre-JWST studies. Therefore, we compute upper limits on the previously missed fraction of the SMF by dividing our SMF by the sum of our SMF with the pre-JWST SMF from \citet{Weaver2023_smf}.

The left panel of Figure \ref{fig:smf_massive_zbins} shows our SMF at $3<z<4$ in comparison with \cite{Weaver2023_smf} (at $z\sim 3.0-3.5$) and \cite{McLeod2021} (at $z\sim 3.25$). At all masses shown in this redshift range, the SMF of our sample lies below the pre-JWST SMF from \cite{Weaver2023_smf} \citep[based on COSMOS2020 observations; see][]{Weaver2023_cosmos} and \cite{McLeod2021} (based on HST and ground-based observations).
The $3<z<4$ SMF deviates the most at the low-mass end but comes closest to the pre-JWST study at the high-mass end, suggesting that up to $\sim30$\% of the galaxy population could have been missed in the pre-JWST SMF from \cite{Weaver2023_smf} at $\mathrm{\log M_\star/M_\odot \sim 10.5}$ and up to $\sim20$\% could have been missed at $\mathrm{\log M_\star/M_\odot \sim 11.0}$
-- dusty galaxies detected with JWST therefore make up a sizeable fraction of the galaxy population at the high-mass end, which suggests that galaxies at the high-mass end have been missing from our galaxy census in this epoch. Further, above $\mathrm{\log M_\star/M_\odot \sim 9.5}$, our SMF at $3<z<4$ is significantly higher than the model SMF from \cite{Long2023} (at $z\sim 3.0-3.5$), with the difference being most pronounced at $\mathrm{\log M_\star/M_\odot \sim 10.5}$. Therefore, we could be seeing an emergent population of main-sequence dusty galaxies that are distinct from the widely studied DSFGs, which are typically more strongly star-forming (and which the \citet{Long2023} simulation is based on). These results indicate that a significant population of obscured galaxies are prevalent at this redshift range.

\renewcommand{\arraystretch}{1.3}
\begin{table}
\caption{Stellar mass function values of massive and dusty galaxies from $3<z<4$ and $4<z<6$, as shown graphically in the first two panels of Figure \ref{fig:smf_massive_zbins}. Uncertainties are calculated as the quadrature addition of Poissonian noise, cosmic variance, and scatter due to SED fitting.}
\centering
\label{tab:smf_data_1}
\begin{tabular}{c|cc}
$\mathrm{\log  M_\star/M_\odot }$ & \multicolumn{2}{c}{$\mathrm{\Phi\ /\ 10^{-5}\ Mpc^{-3} dex^{-1}}$}  \\
 & $3<z<4$ & $4<z<6$ \\
\hline
\hline

9.5 & $4.10_{-2.37}^{+2.62}$ & $2.39_{-1.38}^{+1.53}$ \\
10.0 & $14.74_{-5.95}^{+4.26}$ & $3.35_{-1.73}^{+1.64}$ \\
10.5 & $17.21_{-6.83}^{+4.82}$ & $1.91_{-1.23}^{+1.45}$ \\
11.0 & $3.28_{-2.61}^{+2.22}$ & $0.48_{-0.43}^{+1.02}$ \\
11.5 & <1.31 & <0.77 \\

\end{tabular} 
\end{table}

\renewcommand{\arraystretch}{1.3}
\begin{table}
\caption{Stellar mass function values of massive and dusty galaxies from $6<z<8$, as shown graphically in the third panel of Figure \ref{fig:smf_massive_zbins}. Uncertainties are calculated as the quadrature addition of Poissonian noise, cosmic variance, and scatter due to SED-fitting.}
\centering
\label{tab:smf_data_2}
\begin{tabular}{c|c}
$\mathrm{\log  M_\star/M_\odot }$ & $\mathrm{\Phi\ /\ 10^{-5}\ Mpc^{-3} dex^{-1}}$  \\
 & $6<z<8$ \\
\hline
\hline

9.625 & $1.19_{-0.87}^{+0.99}$ \\
10.375 & $2.38_{-1.35}^{+1.11}$ \\
11.125 & <0.64 \\
 
\end{tabular} 
\end{table}

At $4<z<6$ (central panel of Figure \ref{fig:smf_massive_zbins}), we compare the SMF of our sample to \cite{Weaver2023_smf} (at $z\sim 4.5-5.5$). For reference, the \cite{Stefanon2017b} mass function for LBGs at $z=5$ is shown.
Of particular interest in this epoch is the comparison of our sample to the pre-JWST SMF at the high-mass end, where at $\mathrm{\log M_\star/M_\odot \sim 10.5}$ and 11.0, we find that up to $\sim25\%$ of the SMF could have been missed in the \cite{Weaver2023_smf} mass function.
In addition, like at $3<z<4$, we find an SMF at $4<z<6$ which is significantly higher than the dust-obscured model SMF predicted by the \cite{Long2023} simulation at $z\sim 4.0-6.0$.

At $6<z<8$ (right panel of Figure \ref{fig:smf_massive_zbins}), we compare our SMF to \cite{Weaver2023_smf} (at $z\sim 6.5-7.5$). For reference, the \cite{Stefanon2021} mass function at $z=7$ is shown. Above $\mathrm{\log M_\star/M_\odot \sim 10.0}$, our sample overtakes the \cite{Weaver2023_smf} SMF and exceeds it at $\mathrm{\log M_\star/M_\odot = 10.375}$. This suggests the emergence of an extensive population of galaxies in the Epoch of Reionisation, hidden in the pre-JWST era but constituting a dominant part of the high-mass population.

We see a strong evolution in our SMFs around $z\sim4$ between masses of $9.5 \lesssim \mathrm{\log M_\star/M_\odot \lesssim 11.0}$. Comparing the SMFs at $4<z<6$ and $3<z<4$ in Table \ref{tab:smf_data_1}, we see an increase by a factor of $\sim4$ at $\mathrm{\log M_\star/M_\odot \sim 10.0}$, $\sim9$ at $\mathrm{\log M_\star/M_\odot \sim 10.5}$, and $\sim7$ at $\mathrm{\log M_\star/M_\odot \sim 11.0}$ -- this shows an accelerated evolution in the knee of the SMF at this epoch, suggesting the onset of rapid dust-obscured stellar mass growth at $z\sim4$.

The SMFs of our sample between $4<z<6$ and $6<z<8$ show little evolution. At the high-mass end, we do not see a strong evolution across the whole redshift range, but we are heavily limited by small sample statistics, systematic uncertainties and cosmic variance, making it challenging to comment on SMF properties without a larger sample.

Globally, our analysis of red, dust-obscured galaxies shows that these sources recover a sizeable fraction of the high-mass end of the pre-JWST SMFs from \citet{Weaver2023_smf}. Not only does this reveal the nature of the massive galaxy population, it highlights the efficiency of JWST in characterising the massive end of the galaxy SMF.

\subsection{Integrated stellar mass density}
\label{subsec:integrated_smd}

The cosmic stellar mass density (SMD) is an efficient measure of stellar mass assembly. The total SMD is tightly coupled with the cosmic star-formation rate history, and thus could provide insights into early galaxy build-up such as previous epochs of star-formation and the stellar IMF of early stellar populations \citep{Dickinson2003}. Multiple works have observationally tracked the evolution of the SMD \citep{Stark2009,Gonzalez2010,Davidzon2017,McLeod2021,Weaver2023_smf}, reaching up to $z\sim8-10$ \citep[e.g.,][]{Weaver2023_smf,Stefanon2021}. The observationally determined SMD, however, can be substantially affected if a significant population of high-mass galaxies have been missing in previous observations. This work in part aims to determine the fraction by which pre-JWST studies have underestimated the SMD.

We integrate the measured SMFs presented in Section \ref{subsec:smf_final} in order to get an estimate of the SMD for our galaxy sample. For each redshift bin `$i$', we numerically integrate over the mass bins indexed by `$j$' following Equation \ref{eq:smd}:
\begin{equation}
\label{eq:smd}
    \rho_{i} = \sum_{j={\rm M_{min}}}^{\rm M_{max}} \Phi_{i,j} \cdot {\rm M}_{j} \cdot \Delta \log {\rm M},
\end{equation}
where $\Phi_{i,j}$ is the SMF value inferred via Equation \ref{eq:smf}, ${\rm M}_{j}$ is the the central mass within each mass bin, $\Delta \log {\rm M}$ is the fixed mass bin size, and the limits are given by the mass range covered by our SMFs. Uncertainties on $\rho_{i}$ are calculated via addition in quadrature, where the upper limits in the SMFs contribute to the upper uncertainty on $\rho_{i}$.

We find that the SMD in units of ${\rm [10^5\ M_\odot Mpc^{-3}]}$ is ${\rm 51.6_{-17.2}^{24.8}}$ at $3<z<4$, ${\rm 7.5_{-3.0}^{+13.4}}$ at $4<z<6$ and ${\rm 4.6_{-2.4}^{+6.7}}$ at $6<z<8$. The large uncertainty estimates reflect the uncertainty in the stellar mass functions where we are limited by sample size, especially at the high-mass end. Additionally, the upper limits in the highest mass bins make sizeable contributions to the upper uncertainties on the SMD.

In order to determine the missed SMD fraction in pre-JWST studies at the high-mass end, we compare our results with the \cite{Weaver2023_smf} study. As similarly explained in Section \ref{subsec:smf_final}, we calculate an upper limit on the missed SMD fraction as the SMD of our sample divided by the sum of our SMD and the \citet{Weaver2023_smf} observed SMD, based on the assumption that our two studies together form a more complete survey than pre-JWST studies alone.

We integrate the observed SMFs from \cite{Weaver2023_smf} (shown in Figure \ref{fig:smf_massive_zbins}) in order to estimate the observed pre-JWST total SMD. Given that the \cite{Weaver2023_smf} SMFs do not reach the lower mass limit of our study at $4<z<6$ and $6<z<8$, we expand the Weaver+23 SMFs to lower masses with their Schechter fits down to $\mathrm{\log M_\star/M_\odot=9.25}$, so as to perform a mass-consistent comparison with our sample. We find missed SMD fractions of $19_{-6}^{+9}$\% at $3<z<4$ and $15_{-6}^{+26}$\% at $4<z<6$. At $6<z<8$, we find a missed fraction of $46_{-24}^{+66}$\%, possibly doubling the SMD at this epoch.
Therefore, our results indicate that the SMD could have been underestimated in pre-JWST studies, in particular significantly at $z>6$.
In future studies, it will be imperative to include dust-obscured galaxies at the high-mass end in order to accurately trace stellar mass build-up in the early universe.

\section{Discussion}
\label{sec:discussion}

In this section, we discuss the results of our work in the context of similar studies conducted with JWST's first year of observations on dusty galaxies. We additionally discuss the abundance of massive galaxies that is suggested by our dust-obscured SMFs, compare our SMD estimates with the SMDs estimated from integrating the \citet{Weaver2023_smf} Schechter functions, discuss the move towards redder selection functions, and place this in the context of past work and future studies on galaxy censuses.

\subsection{Comparison of sample to recent literature in CEERS}
\label{subsec:compare_to_lit}

JWST's pilot year has seen the output of a great amount of science, with several papers and teams already providing novel insights into obscured galaxies at $z>3$ \citep[e.g.,][]{Barrufet2023, Nelson2023_pub, Gonzalez2023, Rodighiero2023, Labbe2023a, Akins2023_pub}. Additionally, it was shown that very dusty galaxies can sometimes contaminate extremely high redshift selections \citep[e.g.,][]{Naidu2022,Zavala2023,ArrabalHaro2023_pub}. Here, we discuss our sample in comparison with some select studies in the CEERS field: \citet{Barrufet2023}, \citet{Gonzalez2023}, \citet{Labbe2023a}, and \citet{Naidu2022}.

\citet{Barrufet2023} studied HST-dark galaxies in the CEERS field, identifying massive, obscured galaxies at $z>3$ and into the Epoch of Reionisation. Of the 30 HST-dark sources in their study, we identify 12 in our sample, likely due to the different colour selection. Our SMF results support the findings of \citet{Barrufet2023} that suggest that a significant fraction of massive, obscured sources were previously missing from our galaxy census at $z>3$.

\citet{Gonzalez2023} studied HST-dark and -faint galaxies in the first four NIRCam pointings of the CEERS field, using a selection based on F150W-F356W colours. Out of their sample of 138 HST-dark galaxies, we identify 65 sources in our sample.
Comparing their total sample to our study, we find similar redshift ranges ($\langle z \rangle =\mathrm{3.68_{-1.00}^{+1.60}}$ in their study, $\langle z \rangle =\mathrm{3.46_{-1.35}^{+2.04}}$ in ours) and stellar masses ($\mathrm{\langle \log M_\star / M_\odot \rangle = 10.20_{-0.73}^{+0.46}}$ in their study, $\mathrm{\langle \log M_\star / M_\odot \rangle =  10.15_{-0.50}^{+0.43}}$ in ours). We note however that our redshift distribution has a longer high-end tail, where we find more sources at $z\gtrsim6$ than the \citet{Gonzalez2023} study. This is most likely because we use the longer wavelength F444W filter in our colour selection, where we are possibly picking up the [OIII] line at $z\sim7$.

Using a selection based on blue rest-UV and red rest-optical colours, \citet{Labbe2023a} found six massive galaxies ($\mathrm{M_\star / M_\odot > 10^{10}}$) at $7.4<z<9.1$. We identify two of their sources in our sample (IDs 48444 and 67066). We most likely do not select the remaining four sources in \citet{Labbe2023a} due to their blue rest-UV colour selection. Additionally, one of the \citet{Labbe2023a} sources originally identified as a massive galaxy at $z=8.13$ has now been spectroscopically determined to be a likely AGN candidate at $z=5.64$ \citep{Kocevski2023_pub}; we do not find this source in our sample.

\citet{Naidu2022} proposed a luminous candidate $z\approx17$ or $z\approx5$ galaxy, dubbed ``Schrodinger’s Galaxy'', now confirmed to be an obscured source at $z=4.912\pm0.001$ \citep{ArrabalHaro2023_pub}. We find this galaxy in our sample (ID 81918) at $z = 4.79_{-0.08}^{+0.05}$ with a dust attenuation of $\mathrm{A_V = 1.74_{-0.17}^{+0.11}\ mag}$. Such studies show that there is increasing evidence for a population of massive, obscured galaxies at high redshifts, close to and into the Epoch of Reionisation \citep[see also][]{Fudamoto2021}.

\subsection{Abundance of red galaxies at the high-mass end of SMFs}
\label{subsec:massive_galaxies_abundance}

The SMFs of JWST-detected dust-obscured galaxies in our study point toward an abundance of galaxies at the massive end of the pre-JWST SMF, possibly leading to an excess of the galaxy population with respect to the pre-JWST determined SMF. This abundance is even more pronounced with respect to the Schechter fits from \citet{Weaver2023_smf} (solid lines in Figure \ref{fig:smf_massive_zbins}). Comparing our SMD estimates to those found by integrating the Schechter fits from \citet{Weaver2023_smf} down to $\mathrm{\log M_\star/M_\odot=9.25}$, we find missed SMD fractions of $19_{-6}^{+9}$\% at $3<z<4$ and $18_{-7}^{+32}$\% at $4<z<6$. At $6<z<8$, we find a missed SMD fraction of $52_{-27}^{+76}$\%, effectively doubling the SMD at this epoch. This excess with respect to the Schechter fit at the massive end of the SMFs at $z\sim 3-5$ was shown in \citet{Weaver2023_smf} with a sample of 2${\rm \mu m}$-selected sources from the COSMOS2020 dataset, with hints that this population could be star-forming, dusty galaxies.

It is evident from past work that selecting sources deeper into the NIR results in stronger constraints on the high-mass end of SMFs. At its time, the \citet{Weaver2023_smf} study represented some of the reddest SMFs in comparison with earlier studies \citep[e.g.][]{Davidzon2017, Stefanon2017b, Stefanon2021}. The effect of this is evident from Figure \ref{fig:smf_massive_zbins}, where the \citet{Weaver2023_smf} SMF overtakes the LBG-based \citet{Stefanon2017b} SMF at $4<z<6$ and the \citet{Stefanon2021} SMF at $6<z<8$ at the high mass end. Now, with JWST/NIRCam allowing us to move even deeper into the NIR regime, our study represents the natural next step in the move towards more complete selections: with sources selected based on their 1.5${\rm \mu m}$-4.44${\rm \mu m}$ colour, our study enables a more complete characterisation of massive and dusty galaxies than was possible with previous SMF studies.

Our results reinforce the conclusion that dust-obscured galaxies contribute significantly to the high-mass end of the SMF. In future galaxy censuses with JWST, it will be critical to explore how the SMF measurements at the high-mass end compare with the Schechter formalism of our description of galaxy evolution.

\section{Summary and conclusion}
\label{sec:summary_conclusion}

In this work, we used data from the JWST/CEERS survey \citep{Finkelstein2022b,Finkelstein2023} in the CANDELS/EGS field to identify red, optically-faint galaxies at high redshifts in order to determine the obscured stellar mass function at various epochs in the first two billion years of the history of the universe. Some key results are summarised in the following:

\begin{itemize}
\item Using a colour criterion designed to select red, optically-faint galaxies, we show that we efficiently select massive and dusty galaxies ($\mathrm{\langle \log M_\star/M_\odot \rangle = 10.15_{-0.50}^{+0.43}}$ and $\mathrm{\langle A_V \rangle = 2.71_{-0.91}^{+0.88}\ mag}$) with a majority lying at $z>3$ (see Figures \ref{fig:redshift_hist} and \ref{fig:red_gal_properties_BAGPIPES}).

\item Our sample contains predominantly star-forming galaxies, largely lying on the star-forming main-sequence. They therefore represent a normal population of galaxies without extreme starburst properties (see Figures \ref{fig:UVJ} and \ref{fig:red_gal_sf_ms}). Our sample overlaps with the \cite{Wang2019} sample at the high-mass end and the \cite{Barrufet2023} sample at the low-mass end, showing that our sample of red galaxies has similar star-forming properties to that of HST-dark galaxies (see Figure \ref{fig:red_gal_sf_ms_compare_to_lit}).

\item Our analysis of the obscured galaxy SMF (see Figure \ref{fig:smf_massive_zbins}) shows that in the pre-JWST era, we have missed a significant fraction of galaxies, particularly at the high-mass end of the SMF at redshifts of $z>3$.
The SMFs of red, optically-faint galaxies suggest a missed fraction of $\gtrsim20$\% of the galaxy population in the $3<z<4$ and $4<z<6$ epochs (at $\log M_\star/M_\odot \ge 10.5$).
At $6<z<8$, our SMF overtakes the pre-JWST SMF from \citet{Weaver2023_smf} around $\mathrm{\log M_\star/M_\odot \sim 10.375}$.

\item Our results at $6<z<8$ highlight the importance of accounting for massive, dust-obscured galaxies in the final stages of the Epoch of Reionisation.

\item Our SMFs show a strong evolution at $z\sim4$ at masses of $9.5 \lesssim \mathrm{\log M_\star/M_\odot \lesssim 11.0}$, suggesting the onset of rapid dust-obscured stellar mass assembly in this epoch.

\item The derived stellar mass density of our sources at $\mathrm{\log M_\star/M_\odot \geq 9.25}$ suggests that the missed SMD fraction could be a factor of $\sim$15-20\% at $z\sim3-6$. We find a missed fraction of $\sim$45\% at $z\sim6-8$, possibly doubling the SMD at this epoch.
\end{itemize}

These findings point towards an emergent population of massive, obscured galaxies from $z\sim3$ up to and into the Epoch of Reionisation, supporting the findings of early JWST studies \citep[e.g.,][]{Barrufet2023,Labbe2023a,Akins2023_pub}. The strong evolution of the SMF at $z\sim4$ suggests that this is a period of rapid stellar mass growth in obscured galaxies. Interestingly, $z\sim4$ is also roughly when the obscured SFRD is thought to overtake the un-obscured SFRD, dominating the cosmic star-formation history at later epochs \citep[e.g.,][]{Zavala2021, Bouwens2020, Bouwens2021}.

Our results indicate that obscured stellar mass assembly occurred as early as $z\sim8$, suggesting that the build-up of dusty galaxies could begin close to 600 Myrs after the Big Bang. To further explore the beginning of obscured stellar mass assembly and push the observable redshift boundary farther back, studying the SMF by collating all public JWST surveys is critical. Including surveys such as COSMOS-Web \citep{Casey2023}, PRIMER \citep{Dunlop2021}, UNCOVER \citep{Bezanson2022} and PANORAMIC \citep{Williams2021_jwstprop} will satisfy the need of the hour: larger sample sizes. 
These surveys, and others to come with JWST, will surely result in us establishing a complete census of the massive, dust-obscured galaxy population in the early universe.

\section*{Acknowledgements}

The work presented in this paper is based on observations made with the NASA/ESA/CSA James Webb Space Telescope. The data were obtained from the Mikulski Archive for Space Telescopes at the Space Telescope Science Institute, which is operated by the Association of Universities for Research in Astronomy, Inc., under NASA contract NAS 5-03127 for JWST. These observations are associated with program \#1345.

This work has received funding from the Swiss State Secretariat for Education, Research and Innovation (SERI) under contract number MB22.00072, as well as from the Swiss National Science Foundation (SNSF) through project grant 200020\_207349.
The Cosmic Dawn Center (DAWN) is funded by the Danish National Research Foundation under grant No.\ 140.

We thank the anonymous reviewer for their constructive feedback and close reading of the manuscript which helped improve the clarity of this paper.
We thank Rui Marques-Chaves, Ivan Kramarenko and Damien Korber for useful discussions that helped improve the quality of this work.
RG gratefully acknowledges support from the Inlaks Shivdasani Foundation.
YF acknowledges support from NAOJ ALMA Scientific Research Grant number 2020-16B and support by JSPS KAKENHI Grant Number JP23K13149.
VG gratefully acknowledges support by the ANID BASAL project FB210003 and from ANID FONDECYT Regular 1221310.
For RPN, support for this work was provided by NASA through the NASA Hubble Fellowship grant HST-HF2-51515.001-A awarded by the Space Telescope Science Institute, which is operated by the Association of Universities for Research in Astronomy, Incorporated, under NASA contract NAS5-26555.
MS acknowledges support from the CIDEGENT/2021/059 grant, from project PID2019-109592GB-I00/AEI/10.13039/501100011033 from the Spanish Ministerio de Ciencia e Innovaci\'on - Agencia Estatal de Investigaci\'on. MS also acknowledges the financial support from the MCIN with funding from the European Union NextGenerationEU and Generalitat Valenciana in the call Programa de Planes Complementarios de I+D+i (PRTR 2022) Project (VAL-JPAS), reference ASFAE/2022/025.
The work of CCW is supported by NOIRLab, which is managed by the Association of Universities for Research in Astronomy (AURA) under a cooperative agreement with the National Science Foundation.

Telescope facilities: JWST (NIRCam), HST (ACS and WFC3)

Several publicly available softwares have facilitated this work. We extend our thanks to the authors of the following softwares:
{\tt IPython} \citep{Perez2007},
{\tt jupyter} \citep{jupyter},
{\tt astropy} \citep{Astropy2013,Astropy2018},
{\tt matplotlib} \citep{Hunter2007},
{\tt numpy} \citep{numpy},
{\tt photutils} \citep{Bradley2022},
{\tt scipy} \citep{Virtanen2020},
{\tt EAZY} \citep{Brammer2008},
{\tt BAGPIPES} \citep{Carnall2018},
{\tt GalfitM} \citep{Haussler2013,Vika2015},
{\tt grizli} \citep{Brammer2018},
{\tt SExtractor} \citep{Bertin1996},
{\tt pypher} \citep{Boucaud2016},
{\tt extinction} \citep{Fitzpatrick2007}, 
{\tt GLACiAR2} \citep{Carrasco2018,Leethochawalit2022}.

\section*{Data Availability}

The JWST and HST raw data products used in this work are available via the Mikulski Archive for Space Telescopes (\url{https://mast.stsci.edu}). The combined mosaics are available on github (\url{https://github.com/gbrammer/grizli/blob/master/docs/grizli/image-release-v6.rst}). Additional data presented in this work will be made available by the authors upon request.



\bibliographystyle{mnras}
\bibliography{ms} 



\clearpage

\appendix
\onecolumn
\makeatletter
\renewcommand{\fps@figure}{htbp}
\renewcommand{\fps@table}{htbp}
\makeatother

\section{AGN identification and effect on the SMF}
\label{sec:agn_appendix}

\begin{figure*}
   \centering
   \includegraphics[width=0.5\textwidth]{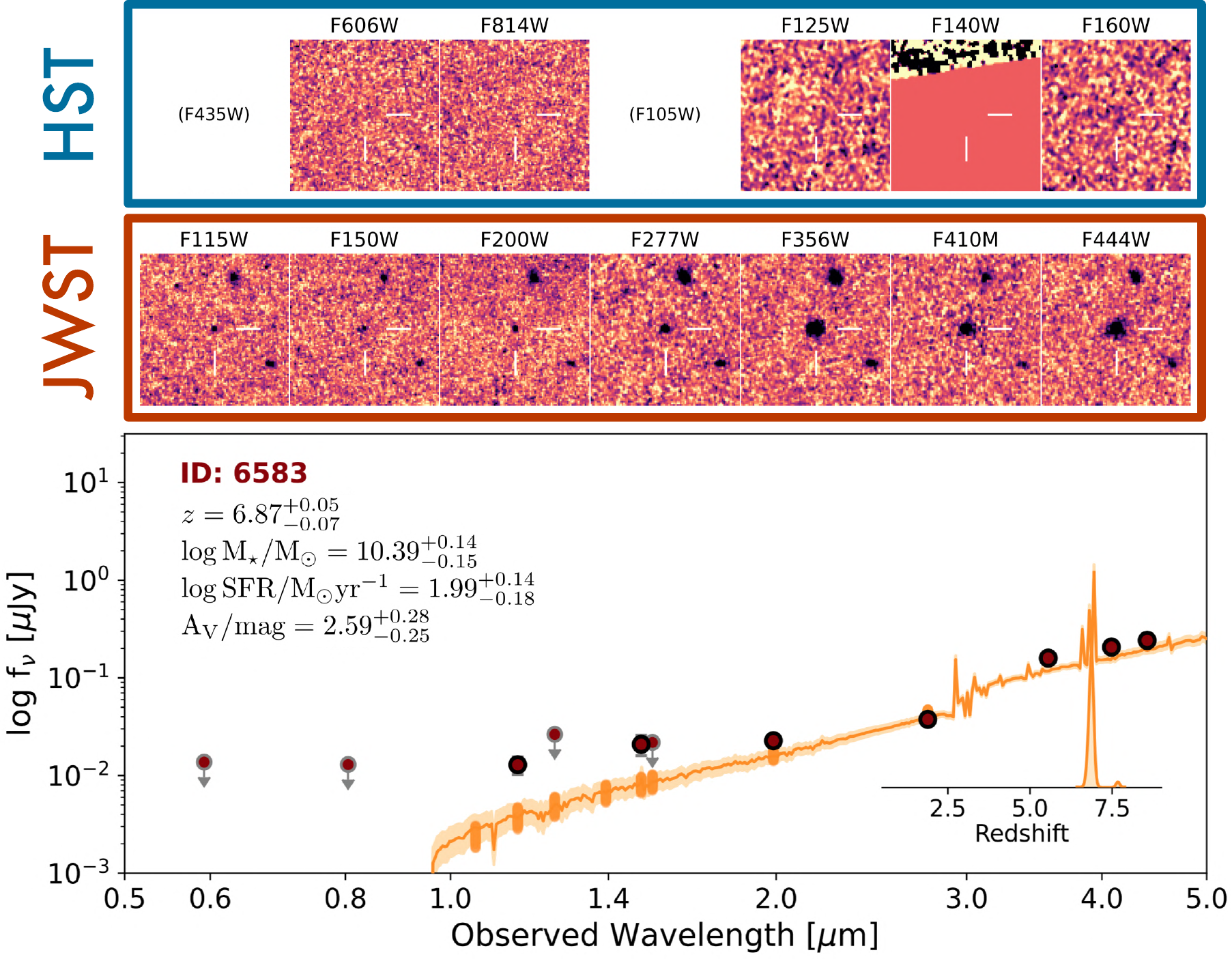}
   \caption{Postage stamps and SED of galaxy 6583, identified as a potential AGN and removed from our final sample. The postage stamps show the compactness of the source. The SED shows a characteristic red slope above 2$\mu$m, and a blue slope below 2$\mu$m. The blue part of the slope is poorly fit with BAGPIPES, thus resulting in an inaccurate photometric redshift and subsequently inaccurate derived physical properties.}
   \label{fig:6583_stamp_n_SED}
\end{figure*}

\begin{figure*}
   \centering
   \includegraphics[width=\textwidth]{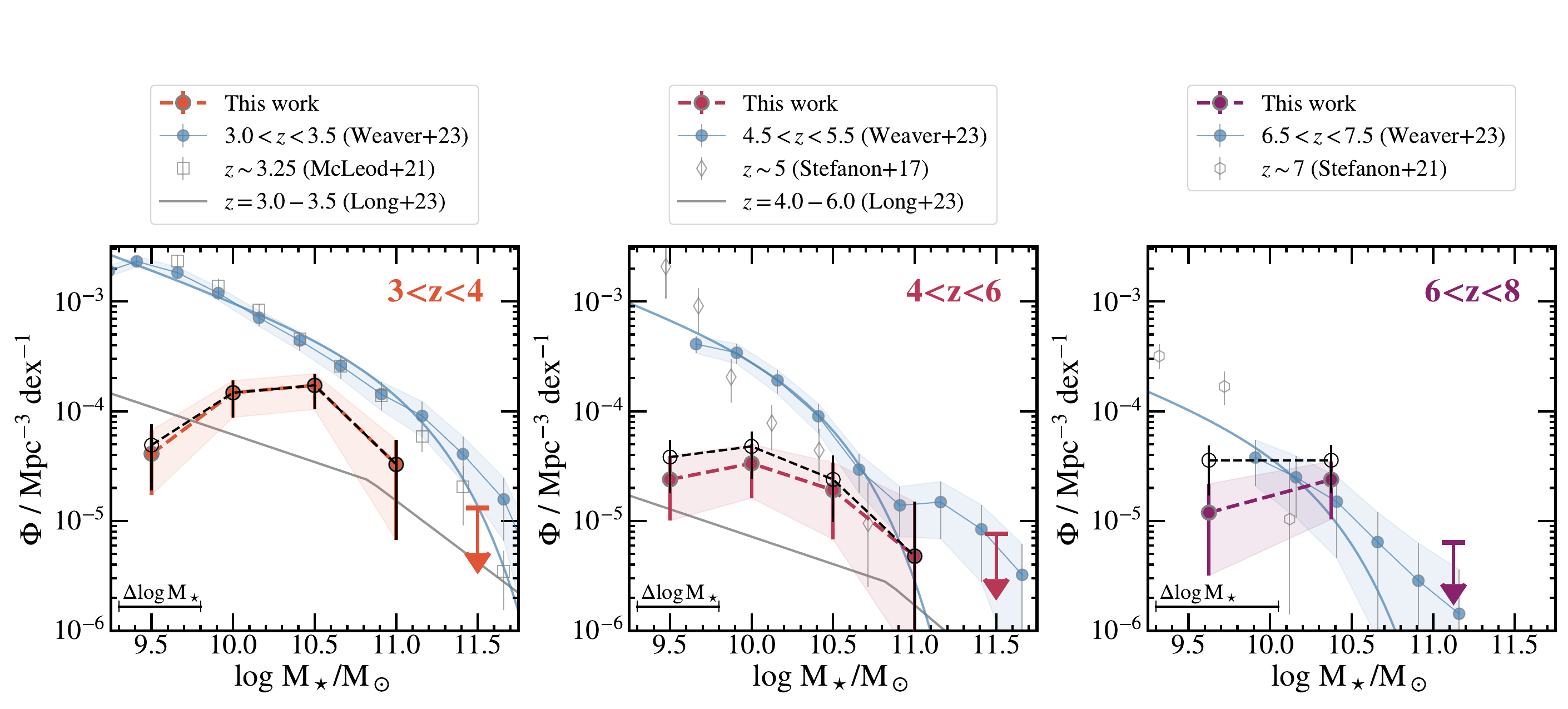}
   \caption{SMFs shown as in Figure \ref{fig:smf_massive_zbins}. In addition, open circles represent the SMFs of the full sample of 168 sources before AGN-removal. The AGN-removal has the least effect on the SMF at $3<z<4$, a larger effect at $4<z<6$ and the most pronounced effect at $6<z<8$. This highlights the importance of identifying and removing AGN candidates at high redshifts in order to derive accurate mass functions and SMD estimates.}
   \label{fig:smf_massive_zbins_full_sample}
\end{figure*}

Given that our study focuses on star-forming galaxies, it is of importance to remove AGN from our sample. We identify and remove AGN candidates, the so-called little red dots (LRDs) as described in Section \ref{subsec:agn}. Figure \ref{fig:6583_stamp_n_SED} shows the postage stamps and SED of one such AGN candidate, galaxy 6583. This is a very compact source, as is characteristic of LRDs, with a red slope beyond 2$\mu$m and a blue slope below this. As shown, BAGPIPES does not fit the short-wavelength end of the slope well, possibly because BAGPIPES cannot perform multi-component SED-fitting and additionally does not contain AGN templates. This results in inaccurate photometric redshifts and derived physical properties of LRDs.
Further, Figure \ref{fig:smf_massive_zbins_full_sample} shows the effect of LRDs on the stellar mass function of our sample. While the SMF of the full sample overlaps neatly with the AGN-cleaned sample at $3<z<4$, the difference between SMFs is more pronounced at $4<z<6$ and differs the most at $6<z<8$. This highlights the importance of addressing the presence of LRDs in our sample, so as not to overestimate the SMFs and stellar mass density at high-redshifts.

\clearpage
\section{Physical properties of galaxies from SED fitting}
\label{sec:tables}

Section \ref{sec:SED_fitting} describes the SED-fitting performed with BAGPIPES. Here, we present the derived physical properties
for the 168 galaxies in our sample: 148 star-forming galaxies and 20 AGN candidates. Table \ref{tab:physical_properties} presents the IDs, RA, Dec, photometric redshifts, stellar masses, SFRs and dust attenuations of all galaxies. AGN candidates are indicated by a $\dagger$.

\begin{table*}
 
\caption{Physical properties of 168 red, optically faint galaxies in our sample. The 20 candidate AGN removed from the final sample are indicated by a $\dagger$.}
\centering
\label{tab:physical_properties}

\begin{tabular}{|c|c|c|c|c|c|c|c|}
\hline
S.No. & ID & RA & Dec & $z_\mathrm{{phot}}$ & $\mathrm{\log\ M_\star/M_\odot}$ & $\mathrm{\log\ SFR/M_\odot yr^{-1}}$ & $\mathrm{A_V\ /mag}$ \\
\hline
\hline

1$^\dagger$        & 2046     &  215.0817208  &  52.9122546   &  7.76$^{+0.30}_{-3.56}$  &  9.79$^{+0.19}_{-0.24}$  &  1.51$^{+0.22}_{-0.40}$  &  2.53$^{+1.22}_{-0.55}$    \\
2        & 2090     &  215.0270820  &  52.8729121   &  1.60$^{+0.06}_{-0.03}$  &  9.83$^{+0.07}_{-0.09}$  &  0.82$^{+0.13}_{-0.13}$  &  2.76$^{+0.22}_{-0.19}$    \\
3        & 2985     &  214.9892637  &  52.8471590   &  0.99$^{+0.07}_{-0.06}$  &  10.00$^{+0.08}_{-0.12}$  &  -2.32$^{+1.67}_{-3.47}$  &  3.52$^{+0.28}_{-0.33}$    \\
4        & 3429     &  215.0884782  &  52.9187800   &  2.01$^{+0.26}_{-0.35}$  &  9.94$^{+0.15}_{-0.18}$  &  1.08$^{+0.25}_{-0.32}$  &  3.33$^{+0.30}_{-0.30}$    \\
5$^\dagger$        & 3812     &  215.1370194  &  52.9556436   &  8.46$^{+0.30}_{-0.32}$  &  9.98$^{+0.21}_{-0.19}$  &  2.00$^{+0.20}_{-0.16}$  &  2.82$^{+0.37}_{-0.33}$    \\
6        & 5145     &  215.1290424  &  52.9518497   &  4.09$^{+0.25}_{-0.37}$  &  9.18$^{+0.15}_{-0.30}$  &  0.82$^{+0.18}_{-0.40}$  &  2.09$^{+0.48}_{-0.31}$    \\
7$^\dagger$        & 6583     &  214.8964766  &  52.7876910   &  6.87$^{+0.05}_{-0.07}$  &  10.39$^{+0.14}_{-0.15}$  &  1.99$^{+0.14}_{-0.18}$  &  2.59$^{+0.28}_{-0.25}$    \\
8        & 6887     &  214.9056138  &  52.7912156   &  2.86$^{+0.09}_{-0.10}$  &  11.32$^{+0.06}_{-0.06}$  &  2.55$^{+0.11}_{-0.09}$  &  3.87$^{+0.09}_{-0.12}$    \\
9        & 7125     &  215.0128418  &  52.8709202   &  2.75$^{+0.12}_{-0.18}$  &  10.71$^{+0.03}_{-0.05}$  &  0.48$^{+0.39}_{-0.63}$  &  1.82$^{+0.25}_{-0.27}$    \\
10       & 7948     &  215.0804330  &  52.9215508   &  3.02$^{+0.47}_{-1.34}$  &  10.28$^{+0.12}_{-0.30}$  &  1.60$^{+0.19}_{-1.68}$  &  3.11$^{+0.50}_{-0.32}$    \\
11       & 8099     &  214.9806159  &  52.8486131   &  2.10$^{+0.11}_{-0.11}$  &  10.43$^{+0.07}_{-0.11}$  &  1.59$^{+0.19}_{-0.12}$  &  3.45$^{+0.22}_{-0.22}$    \\
12       & 8730     &  214.8544200  &  52.7595804   &  1.92$^{+0.08}_{-0.06}$  &  10.31$^{+0.05}_{-0.10}$  &  1.40$^{+0.16}_{-0.10}$  &  3.12$^{+0.20}_{-0.19}$    \\
13       & 9030     &  215.0445212  &  52.8971830   &  4.53$^{+0.12}_{-0.14}$  &  9.76$^{+0.11}_{-0.08}$  &  1.66$^{+0.07}_{-0.13}$  &  1.85$^{+0.14}_{-0.16}$    \\
14       & 9646     &  215.0759978  &  52.9213333   &  4.81$^{+0.10}_{-0.16}$  &  9.86$^{+0.06}_{-0.05}$  &  1.29$^{+0.09}_{-0.08}$  &  1.60$^{+0.12}_{-0.13}$    \\
15       & 9823     &  214.8970451  &  52.7922215   &  2.99$^{+0.06}_{-0.08}$  &  10.41$^{+0.05}_{-0.11}$  &  1.72$^{+0.17}_{-0.10}$  &  2.24$^{+0.19}_{-0.14}$    \\
16       & 10083    &  215.0348553  &  52.8913610   &  2.70$^{+0.11}_{-0.16}$  &  11.06$^{+0.05}_{-0.06}$  &  2.26$^{+0.12}_{-0.13}$  &  3.79$^{+0.14}_{-0.15}$    \\
17       & 11399    &  214.9934658  &  52.8643256   &  2.28$^{+0.39}_{-0.11}$  &  10.29$^{+0.08}_{-0.10}$  &  1.26$^{+0.22}_{-0.61}$  &  2.75$^{+0.46}_{-0.54}$    \\
18       & 12620    &  214.8645568  &  52.7742302   &  4.13$^{+0.18}_{-2.30}$  &  9.33$^{+0.09}_{-0.56}$  &  0.82$^{+0.15}_{-0.74}$  &  2.10$^{+0.94}_{-0.21}$    \\
19       & 13192    &  214.9117928  &  52.8090634   &  6.33$^{+0.12}_{-0.11}$  &  10.27$^{+0.05}_{-0.05}$  &  1.14$^{+0.18}_{-0.16}$  &  1.00$^{+0.18}_{-0.16}$    \\
20       & 13632    &  214.8998108  &  52.8015440   &  8.90$^{+0.76}_{-0.29}$  &  9.83$^{+0.29}_{-0.21}$  &  1.61$^{+0.31}_{-0.20}$  &  1.93$^{+0.31}_{-0.25}$    \\
21       & 13789    &  215.1506541  &  52.9801120   &  1.89$^{+0.10}_{-0.07}$  &  10.18$^{+0.07}_{-0.09}$  &  1.23$^{+0.14}_{-0.14}$  &  2.89$^{+0.22}_{-0.17}$    \\
22       & 14669    &  215.0722340  &  52.9253267   &  5.23$^{+0.06}_{-0.11}$  &  10.28$^{+0.09}_{-0.09}$  &  2.12$^{+0.06}_{-0.08}$  &  2.15$^{+0.10}_{-0.09}$    \\
23$^\dagger$       & 14807    &  214.9551930  &  52.8430203   &  5.37$^{+0.04}_{-0.03}$  &  9.71$^{+0.13}_{-0.11}$  &  1.72$^{+0.06}_{-0.07}$  &  2.20$^{+0.10}_{-0.11}$    \\
24$^\dagger$       & 15203    &  214.9349717  &  52.8293673   &  5.97$^{+0.31}_{-0.18}$  &  9.23$^{+0.14}_{-0.10}$  &  1.25$^{+0.09}_{-0.10}$  &  2.05$^{+0.19}_{-0.22}$    \\
25       & 15328    &  214.9799606  &  52.8610729   &  6.94$^{+0.12}_{-0.06}$  &  10.17$^{+0.04}_{-0.05}$  &  1.26$^{+0.08}_{-0.06}$  &  0.52$^{+0.09}_{-0.05}$    \\
26       & 15607    &  214.9711828  &  52.8548811   &  3.03$^{+0.03}_{-0.03}$  &  9.93$^{+0.03}_{-0.02}$  &  2.02$^{+0.02}_{-0.02}$  &  2.85$^{+0.04}_{-0.05}$    \\
27       & 15791    &  214.9438346  &  52.8358144   &  5.44$^{+0.05}_{-0.06}$  &  10.52$^{+0.08}_{-0.09}$  &  2.08$^{+0.09}_{-0.08}$  &  2.32$^{+0.12}_{-0.10}$    \\
28       & 15973    &  214.9831267  &  52.8639966   &  2.74$^{+0.07}_{-0.09}$  &  10.36$^{+0.06}_{-0.08}$  &  1.62$^{+0.12}_{-0.13}$  &  2.23$^{+0.16}_{-0.13}$    \\
29       & 16514    &  214.8097475  &  52.7396845   &  2.19$^{+0.17}_{-0.12}$  &  10.27$^{+0.10}_{-0.12}$  &  1.34$^{+0.15}_{-0.23}$  &  3.30$^{+0.27}_{-0.44}$    \\
30       & 18000    &  214.8540675  &  52.7735451   &  3.14$^{+0.16}_{-0.48}$  &  10.23$^{+0.10}_{-0.32}$  &  1.66$^{+0.12}_{-0.14}$  &  2.22$^{+0.50}_{-0.24}$    \\
31       & 18027    &  214.8528902  &  52.7739402   &  3.91$^{+0.15}_{-0.11}$  &  10.29$^{+0.06}_{-0.12}$  &  1.73$^{+0.14}_{-0.12}$  &  2.00$^{+0.19}_{-0.16}$    \\
32       & 18652    &  214.8250625  &  52.7557775   &  3.24$^{+0.37}_{-0.45}$  &  8.79$^{+0.08}_{-0.15}$  &  0.07$^{+0.14}_{-0.20}$  &  1.41$^{+0.35}_{-0.18}$    \\
33       & 19364    &  214.8906538  &  52.8030515   &  5.13$^{+0.13}_{-1.42}$  &  11.04$^{+0.10}_{-0.33}$  &  2.56$^{+0.09}_{-0.61}$  &  3.84$^{+0.13}_{-0.19}$    \\
34       & 19829    &  214.8829253  &  52.7981532   &  3.47$^{+0.09}_{-0.13}$  &  10.56$^{+0.07}_{-0.10}$  &  1.92$^{+0.12}_{-0.10}$  &  3.77$^{+0.15}_{-0.17}$    \\
35       & 20407    &  214.9966641  &  52.8805028   &  2.33$^{+0.03}_{-0.03}$  &  9.88$^{+0.15}_{-0.12}$  &  1.80$^{+0.04}_{-0.08}$  &  3.13$^{+0.12}_{-0.16}$    \\
36$^\dagger$       & 21236    &  214.8997094  &  52.8128429   &  5.46$^{+1.53}_{-0.05}$  &  9.80$^{+0.29}_{-0.17}$  &  1.75$^{+0.11}_{-0.08}$  &  2.48$^{+0.17}_{-0.48}$    \\
37       & 21274    &  214.9613710  &  52.8574003   &  5.18$^{+0.05}_{-0.06}$  &  9.71$^{+0.09}_{-0.10}$  &  1.70$^{+0.06}_{-0.05}$  &  2.56$^{+0.16}_{-0.15}$    \\
38       & 21534    &  215.0413182  &  52.9140811   &  2.14$^{+0.53}_{-0.30}$  &  10.75$^{+0.10}_{-0.09}$  &  1.68$^{+0.32}_{-1.80}$  &  3.45$^{+0.40}_{-0.25}$    \\
39       & 21642    &  214.8191823  &  52.7553065   &  3.81$^{+0.11}_{-1.21}$  &  9.65$^{+0.19}_{-0.21}$  &  1.50$^{+0.08}_{-0.36}$  &  1.98$^{+0.22}_{-0.20}$    \\
40$^\dagger$       & 23057    &  214.8945658  &  52.8121655   &  5.77$^{+0.23}_{-0.18}$  &  10.30$^{+0.12}_{-0.18}$  &  2.00$^{+0.14}_{-0.16}$  &  3.02$^{+0.29}_{-0.24}$    \\
41       & 24455    &  214.9331814  &  52.8415583   &  3.49$^{+0.07}_{-0.29}$  &  9.74$^{+0.09}_{-0.11}$  &  1.22$^{+0.14}_{-0.15}$  &  1.61$^{+0.19}_{-0.14}$    \\
42       & 24824    &  215.1370832  &  52.9885559   &  7.31$^{+0.06}_{-0.05}$  &  9.64$^{+0.07}_{-0.06}$  &  1.72$^{+0.06}_{-0.05}$  &  1.90$^{+0.10}_{-0.09}$    \\

\end{tabular} 
\end{table*}

\begin{table*}
 
\contcaption{}
\centering
\label{tab:continued1}

\begin{tabular}{|c|c|c|c|c|c|c|c|}
\hline
S.No. & ID & RA & Dec & $z_\mathrm{{phot}}$ & $\mathrm{\log\ M_\star/M_\odot}$ & $\mathrm{\log\ SFR/M_\odot yr^{-1}}$ & $\mathrm{A_V\ /mag}$ \\
\hline
\hline

43       & 24835    &  214.9076300  &  52.8234531   &  3.56$^{+0.09}_{-0.07}$  &  10.50$^{+0.09}_{-0.14}$  &  1.97$^{+0.13}_{-0.12}$  &  3.85$^{+0.11}_{-0.16}$    \\
44       & 25691    &  215.0154584  &  52.9009634   &  3.60$^{+0.05}_{-0.05}$  &  10.16$^{+0.05}_{-0.04}$  &  2.23$^{+0.03}_{-0.04}$  &  2.63$^{+0.06}_{-0.06}$    \\
45       & 25864    &  214.9967509  &  52.8890337   &  3.59$^{+0.07}_{-0.06}$  &  10.29$^{+0.11}_{-0.09}$  &  2.27$^{+0.04}_{-0.04}$  &  2.73$^{+0.09}_{-0.10}$    \\
46       & 25923    &  215.1357255  &  52.9875677   &  1.49$^{+0.06}_{-0.08}$  &  9.88$^{+0.09}_{-0.12}$  &  0.94$^{+0.14}_{-0.12}$  &  3.05$^{+0.27}_{-0.23}$    \\
47       & 26371    &  214.8748937  &  52.8014448   &  2.85$^{+0.08}_{-0.08}$  &  10.27$^{+0.06}_{-0.12}$  &  1.50$^{+0.17}_{-0.09}$  &  1.78$^{+0.22}_{-0.10}$    \\
48       & 27302    &  214.8949216  &  52.8171584   &  3.01$^{+0.05}_{-0.07}$  &  10.45$^{+0.10}_{-0.19}$  &  1.92$^{+0.21}_{-0.18}$  &  2.92$^{+0.26}_{-0.20}$    \\
49$^\dagger$       & 27673    &  214.8760429  &  52.8061119   &  7.86$^{+0.06}_{-0.06}$  &  9.90$^{+0.07}_{-0.08}$  &  1.61$^{+0.10}_{-0.07}$  &  1.47$^{+0.15}_{-0.11}$    \\
50       & 27813    &  215.0312932  &  52.9171046   &  3.47$^{+0.07}_{-0.09}$  &  10.29$^{+0.10}_{-0.13}$  &  1.85$^{+0.12}_{-0.12}$  &  3.79$^{+0.14}_{-0.22}$    \\
51       & 28768    &  214.9813321  &  52.8825635   &  4.42$^{+0.12}_{-0.12}$  &  10.28$^{+0.07}_{-0.09}$  &  1.88$^{+0.13}_{-0.14}$  &  2.18$^{+0.16}_{-0.16}$    \\
52       & 29661    &  214.8607038  &  52.7968401   &  3.04$^{+0.15}_{-0.05}$  &  10.85$^{+0.07}_{-0.09}$  &  2.15$^{+0.14}_{-0.11}$  &  3.50$^{+0.22}_{-0.19}$    \\
53       & 29799    &  214.7833571  &  52.7418112   &  3.58$^{+0.07}_{-0.05}$  &  9.61$^{+0.08}_{-0.06}$  &  1.63$^{+0.04}_{-0.04}$  &  1.79$^{+0.05}_{-0.08}$    \\
54       & 29820    &  214.9623604  &  52.8700379   &  3.29$^{+0.28}_{-0.34}$  &  10.42$^{+0.23}_{-0.16}$  &  2.19$^{+0.12}_{-0.19}$  &  3.52$^{+0.21}_{-0.23}$    \\
55       & 30392    &  214.8125539  &  52.7627779   &  2.70$^{+1.09}_{-0.18}$  &  9.77$^{+0.19}_{-0.15}$  &  1.08$^{+0.29}_{-0.17}$  &  1.95$^{+0.26}_{-0.27}$    \\
56       & 30545    &  214.8852115  &  52.8157469   &  2.84$^{+0.23}_{-0.16}$  &  10.51$^{+0.15}_{-0.18}$  &  2.14$^{+0.10}_{-0.18}$  &  3.52$^{+0.23}_{-0.27}$    \\
57       & 31590    &  215.0110259  &  52.9080516   &  3.37$^{+0.07}_{-0.08}$  &  10.21$^{+0.05}_{-0.07}$  &  1.54$^{+0.11}_{-0.09}$  &  1.67$^{+0.13}_{-0.10}$    \\
58       & 32459    &  214.8327532  &  52.7813617   &  3.73$^{+0.35}_{-0.11}$  &  10.25$^{+0.14}_{-0.18}$  &  1.82$^{+0.13}_{-0.17}$  &  3.67$^{+0.21}_{-0.26}$    \\
59       & 33202    &  214.8052770  &  52.7628116   &  3.07$^{+0.78}_{-0.52}$  &  9.89$^{+0.20}_{-0.32}$  &  1.44$^{+0.27}_{-0.27}$  &  2.40$^{+0.40}_{-0.29}$    \\
60       & 33383    &  214.7737355  &  52.7403918   &  1.00$^{+0.12}_{-0.09}$  &  8.41$^{+0.19}_{-0.17}$  &  -0.68$^{+0.24}_{-0.23}$  &  3.43$^{+0.36}_{-0.66}$    \\
61$^\dagger$       & 33394    &  214.9241508  &  52.8490510   &  4.83$^{+0.05}_{-0.04}$  &  10.22$^{+0.06}_{-0.06}$  &  1.57$^{+0.08}_{-0.29}$  &  1.78$^{+0.11}_{-0.38}$    \\
62       & 33621    &  214.7739154  &  52.7413502   &  8.04$^{+0.41}_{-0.31}$  &  9.80$^{+0.10}_{-0.19}$  &  1.57$^{+0.10}_{-0.14}$  &  1.68$^{+0.18}_{-0.15}$    \\
63       & 34437    &  214.7738211  &  52.7400098   &  3.50$^{+0.07}_{-0.16}$  &  10.58$^{+0.06}_{-0.10}$  &  1.96$^{+0.13}_{-0.11}$  &  2.22$^{+0.17}_{-0.12}$    \\
64       & 35262    &  214.8464660  &  52.7959697   &  1.86$^{+0.34}_{-0.28}$  &  9.88$^{+0.17}_{-0.16}$  &  0.88$^{+0.36}_{-0.27}$  &  3.54$^{+0.28}_{-0.39}$    \\
65$^\dagger$       & 35580    &  214.8501142  &  52.8000522   &  4.10$^{+0.37}_{-2.53}$  &  9.27$^{+0.10}_{-0.75}$  &  0.64$^{+0.17}_{-0.92}$  &  1.62$^{+1.14}_{-0.22}$    \\
66       & 36882    &  214.8426487  &  52.7954529   &  1.55$^{+0.07}_{-0.09}$  &  9.78$^{+0.09}_{-0.12}$  &  0.83$^{+0.18}_{-0.12}$  &  3.14$^{+0.32}_{-0.26}$    \\
67       & 39594    &  214.7713801  &  52.7497509   &  3.63$^{+0.09}_{-0.09}$  &  10.64$^{+0.08}_{-0.06}$  &  1.91$^{+0.09}_{-0.82}$  &  3.85$^{+0.11}_{-0.44}$    \\
68       & 40635    &  214.8182953  &  52.7863215   &  1.84$^{+0.14}_{-0.19}$  &  10.07$^{+0.10}_{-0.11}$  &  1.10$^{+0.21}_{-0.21}$  &  3.67$^{+0.22}_{-0.25}$    \\
69       & 40641    &  214.8402710  &  52.8011104   &  6.19$^{+0.24}_{-0.28}$  &  10.47$^{+0.11}_{-0.12}$  &  2.30$^{+0.10}_{-0.11}$  &  1.87$^{+0.13}_{-0.16}$    \\
70       & 41002    &  214.8550845  &  52.8130408   &  4.01$^{+0.12}_{-0.10}$  &  9.90$^{+0.05}_{-0.07}$  &  1.23$^{+0.10}_{-0.07}$  &  1.46$^{+0.13}_{-0.09}$    \\
71       & 41028    &  214.9415578  &  52.8742101   &  2.90$^{+0.11}_{-0.08}$  &  10.22$^{+0.11}_{-0.14}$  &  1.75$^{+0.12}_{-0.16}$  &  1.91$^{+0.17}_{-0.19}$    \\
72       & 41343    &  214.7993284  &  52.7740023   &  5.17$^{+0.04}_{-0.05}$  &  9.14$^{+0.08}_{-0.08}$  &  1.15$^{+0.06}_{-0.05}$  &  1.93$^{+0.11}_{-0.13}$    \\
73       & 41769    &  214.8183966  &  52.7863542   &  2.39$^{+0.27}_{-0.27}$  &  10.48$^{+0.13}_{-0.13}$  &  2.15$^{+0.15}_{-0.22}$  &  3.66$^{+0.26}_{-0.22}$    \\
74$^\dagger$       & 42428    &  214.8493875  &  52.8118246   &  6.24$^{+0.21}_{-0.19}$  &  10.10$^{+0.07}_{-0.08}$  &  1.63$^{+0.11}_{-0.09}$  &  1.77$^{+0.11}_{-0.11}$    \\
75$^\dagger$       & 43895    &  214.9909773  &  52.9165225   &  7.88$^{+0.19}_{-1.98}$  &  10.33$^{+0.12}_{-0.12}$  &  2.19$^{+0.14}_{-0.21}$  &  2.38$^{+0.28}_{-0.23}$    \\
76       & 44383    &  214.7610833  &  52.7506849   &  3.80$^{+0.06}_{-0.85}$  &  10.13$^{+0.12}_{-0.33}$  &  1.79$^{+0.14}_{-0.25}$  &  1.77$^{+0.22}_{-0.18}$    \\
77       & 44999    &  214.8967047  &  52.8497952   &  2.12$^{+0.09}_{-0.08}$  &  10.36$^{+0.09}_{-0.14}$  &  1.61$^{+0.18}_{-0.17}$  &  3.12$^{+0.27}_{-0.26}$    \\
78       & 45609    &  214.8871213  &  52.8453774   &  3.65$^{+0.25}_{-0.10}$  &  9.46$^{+0.10}_{-0.17}$  &  0.93$^{+0.12}_{-0.13}$  &  1.80$^{+0.18}_{-0.20}$    \\
79       & 46100    &  214.8098740  &  52.7894326   &  3.52$^{+0.11}_{-0.32}$  &  10.48$^{+0.08}_{-0.15}$  &  1.87$^{+0.13}_{-0.13}$  &  3.03$^{+0.32}_{-0.18}$    \\
80       & 48444    &  214.8405363  &  52.8179423   &  8.14$^{+0.23}_{-0.18}$  &  9.65$^{+0.18}_{-0.13}$  &  1.67$^{+0.14}_{-0.13}$  &  2.10$^{+0.25}_{-0.21}$    \\
81       & 50438    &  214.7352152  &  52.7451418   &  3.13$^{+0.46}_{-0.26}$  &  10.50$^{+0.15}_{-0.13}$  &  1.84$^{+0.19}_{-0.19}$  &  3.72$^{+0.19}_{-0.27}$    \\
82       & 50590    &  214.7338969  &  52.7444469   &  2.15$^{+0.29}_{-0.14}$  &  10.24$^{+0.12}_{-0.14}$  &  1.50$^{+0.17}_{-0.22}$  &  3.08$^{+0.24}_{-0.35}$    \\
83       & 51072    &  214.9295156  &  52.8879151   &  7.26$^{+0.26}_{-0.17}$  &  10.58$^{+0.13}_{-0.14}$  &  2.28$^{+0.12}_{-0.14}$  &  3.66$^{+0.22}_{-0.25}$    \\
84       & 51077    &  214.9785566  &  52.9215403   &  2.51$^{+0.08}_{-0.06}$  &  10.44$^{+0.04}_{-0.05}$  &  -1.32$^{+0.77}_{-1.27}$  &  1.01$^{+0.30}_{-0.17}$    \\
85       & 51978    &  214.8706665  &  52.8461073   &  3.58$^{+0.09}_{-0.07}$  &  10.10$^{+0.14}_{-0.15}$  &  1.72$^{+0.14}_{-0.18}$  &  2.56$^{+0.21}_{-0.19}$    \\
86       & 52049    &  214.7230124  &  52.7397625   &  3.60$^{+0.19}_{-0.11}$  &  10.74$^{+0.10}_{-0.09}$  &  2.06$^{+0.13}_{-0.10}$  &  3.93$^{+0.05}_{-0.14}$    \\
87       & 52288    &  214.8403421  &  52.8249495   &  2.29$^{+0.07}_{-0.07}$  &  10.68$^{+0.05}_{-0.06}$  &  1.81$^{+0.11}_{-0.11}$  &  3.61$^{+0.19}_{-0.21}$    \\
88       & 52954    &  214.7291876  &  52.7446890   &  2.45$^{+0.04}_{-0.05}$  &  10.03$^{+0.10}_{-0.11}$  &  1.85$^{+0.06}_{-0.09}$  &  2.71$^{+0.13}_{-0.10}$    \\
89       & 53395    &  214.8624249  &  52.8429058   &  2.88$^{+0.26}_{-0.23}$  &  10.21$^{+0.18}_{-0.14}$  &  2.06$^{+0.11}_{-0.11}$  &  3.19$^{+0.20}_{-0.18}$    \\
90       & 54147    &  214.8296607  &  52.8207741   &  3.63$^{+0.07}_{-0.06}$  &  10.36$^{+0.14}_{-0.12}$  &  2.09$^{+0.10}_{-0.15}$  &  3.16$^{+0.13}_{-0.17}$    \\

\end{tabular} 
\end{table*}

\begin{table*}
 
\contcaption{}
\centering
\label{tab:continued2}

\begin{tabular}{|c|c|c|c|c|c|c|c|}
\hline
S.No. & ID & RA & Dec & $z_\mathrm{{phot}}$ & $\mathrm{\log\ M_\star/M_\odot}$ & $\mathrm{\log\ SFR/M_\odot yr^{-1}}$ & $\mathrm{A_V\ /mag}$ \\
\hline
\hline

91       & 55631    &  214.8099701  &  52.8097415   &  3.68$^{+0.31}_{-0.09}$  &  10.78$^{+0.27}_{-0.12}$  &  2.70$^{+0.06}_{-0.09}$  &  3.73$^{+0.11}_{-0.22}$    \\
92       & 56217    &  214.9475951  &  52.9111224   &  1.72$^{+0.08}_{-0.08}$  &  9.62$^{+0.11}_{-0.10}$  &  0.55$^{+0.20}_{-0.78}$  &  2.69$^{+0.26}_{-0.60}$    \\
93       & 56832    &  214.8769354  &  52.8603939   &  2.68$^{+0.12}_{-0.33}$  &  10.62$^{+0.08}_{-0.14}$  &  1.88$^{+0.18}_{-0.18}$  &  3.13$^{+0.27}_{-0.19}$    \\
94       & 56856    &  214.9119355  &  52.8857565   &  3.56$^{+0.12}_{-0.05}$  &  9.80$^{+0.09}_{-0.08}$  &  1.79$^{+0.05}_{-0.04}$  &  1.75$^{+0.06}_{-0.08}$    \\
95       & 57143    &  214.8719994  &  52.8593098   &  7.33$^{+0.14}_{-0.09}$  &  8.77$^{+0.15}_{-0.16}$  &  0.76$^{+0.13}_{-0.10}$  &  1.16$^{+0.22}_{-0.19}$    \\
96       & 57734    &  214.7181000  &  52.7481020   &  2.65$^{+0.51}_{-0.15}$  &  10.03$^{+0.20}_{-0.19}$  &  1.43$^{+0.26}_{-0.17}$  &  3.57$^{+0.25}_{-0.38}$    \\
97       & 57837    &  214.8348814  &  52.8323851   &  2.45$^{+0.12}_{-0.08}$  &  10.21$^{+0.10}_{-0.13}$  &  1.63$^{+0.10}_{-0.16}$  &  2.27$^{+0.19}_{-0.20}$    \\
98       & 57866    &  214.9165099  &  52.8907640   &  2.83$^{+0.11}_{-0.12}$  &  10.71$^{+0.07}_{-0.10}$  &  1.97$^{+0.16}_{-0.13}$  &  3.76$^{+0.15}_{-0.17}$    \\
99       & 58375    &  214.9135356  &  52.8910105   &  4.01$^{+0.36}_{-2.21}$  &  9.21$^{+0.21}_{-0.49}$  &  0.84$^{+0.23}_{-0.91}$  &  2.07$^{+0.77}_{-0.32}$    \\
100      & 58801    &  214.7632325  &  52.7826809   &  2.17$^{+1.49}_{-0.40}$  &  10.08$^{+0.35}_{-0.20}$  &  1.21$^{+0.76}_{-0.35}$  &  3.59$^{+0.32}_{-0.79}$    \\
101      & 59223    &  214.8922499  &  52.8774089   &  7.28$^{+0.15}_{-0.13}$  &  10.65$^{+0.04}_{-0.03}$  &  1.68$^{+0.07}_{-0.06}$  &  1.24$^{+0.09}_{-0.06}$    \\
102      & 60533    &  214.8560303  &  52.8546725   &  3.67$^{+0.45}_{-0.32}$  &  9.97$^{+0.29}_{-0.18}$  &  1.77$^{+0.11}_{-0.39}$  &  3.26$^{+0.23}_{-0.43}$    \\
103      & 60809    &  214.8557255  &  52.8546205   &  3.10$^{+0.18}_{-0.13}$  &  9.50$^{+0.07}_{-0.11}$  &  0.79$^{+0.13}_{-0.11}$  &  1.78$^{+0.22}_{-0.19}$    \\
104      & 61017    &  214.8558862  &  52.8546713   &  1.89$^{+1.73}_{-0.22}$  &  9.86$^{+0.42}_{-0.14}$  &  0.94$^{+0.86}_{-0.26}$  &  3.51$^{+0.33}_{-0.91}$    \\
105      & 61155    &  214.9050035  &  52.8903902   &  3.57$^{+0.11}_{-0.08}$  &  10.03$^{+0.17}_{-0.13}$  &  1.83$^{+0.08}_{-0.16}$  &  2.97$^{+0.15}_{-0.14}$    \\
106      & 61732    &  214.9509311  &  52.9239597   &  1.80$^{+0.18}_{-0.34}$  &  9.94$^{+0.14}_{-0.15}$  &  1.05$^{+0.23}_{-0.24}$  &  3.30$^{+0.34}_{-0.28}$    \\
107      & 62882    &  214.7581773  &  52.7872067   &  2.80$^{+0.08}_{-0.07}$  &  10.07$^{+0.06}_{-0.10}$  &  1.33$^{+0.17}_{-0.13}$  &  2.21$^{+0.19}_{-0.16}$    \\
108      & 63309    &  214.8476073  &  52.8534055   &  4.19$^{+0.42}_{-0.59}$  &  9.84$^{+0.15}_{-0.29}$  &  1.35$^{+0.20}_{-0.38}$  &  3.28$^{+0.31}_{-0.25}$    \\
109      & 63467    &  214.8475506  &  52.8533680   &  4.28$^{+0.47}_{-0.39}$  &  10.25$^{+0.18}_{-0.16}$  &  1.65$^{+0.20}_{-0.19}$  &  3.86$^{+0.10}_{-0.19}$    \\
110      & 63642    &  214.8588311  &  52.8603958   &  1.96$^{+0.52}_{-0.40}$  &  10.02$^{+0.21}_{-0.24}$  &  1.17$^{+0.27}_{-0.38}$  &  3.34$^{+0.40}_{-0.48}$    \\
111      & 63912    &  214.9375051  &  52.9182908   &  3.94$^{+0.07}_{-0.07}$  &  9.98$^{+0.10}_{-0.11}$  &  1.68$^{+0.10}_{-0.20}$  &  2.03$^{+0.14}_{-0.18}$    \\
112      & 64408    &  215.0229080  &  52.9800661   &  3.58$^{+0.07}_{-0.04}$  &  9.90$^{+0.10}_{-0.08}$  &  1.90$^{+0.04}_{-0.04}$  &  2.17$^{+0.07}_{-0.09}$    \\
113      & 65999    &  214.7189003  &  52.7643900   &  5.27$^{+0.06}_{-0.22}$  &  9.80$^{+0.08}_{-0.09}$  &  1.38$^{+0.08}_{-0.08}$  &  1.62$^{+0.12}_{-0.11}$    \\
114      & 66597    &  214.9254260  &  52.9133973   &  2.31$^{+0.08}_{-0.21}$  &  9.94$^{+0.11}_{-0.14}$  &  1.23$^{+0.16}_{-0.16}$  &  2.48$^{+0.23}_{-0.25}$    \\
115      & 66608    &  214.8538962  &  52.8613647   &  3.33$^{+0.06}_{-0.09}$  &  11.24$^{+0.06}_{-0.13}$  &  2.62$^{+0.16}_{-0.11}$  &  3.44$^{+0.23}_{-0.15}$    \\
116$^\dagger$      & 66755    &  214.6951558  &  52.7485691   &  9.32$^{+0.42}_{-0.51}$  &  10.55$^{+0.19}_{-0.25}$  &  2.38$^{+0.22}_{-0.29}$  &  2.93$^{+0.32}_{-0.28}$    \\
117      & 66989    &  215.0368171  &  52.9935017   &  1.82$^{+0.09}_{-0.16}$  &  9.82$^{+0.07}_{-0.14}$  &  0.86$^{+0.21}_{-0.22}$  &  2.67$^{+0.23}_{-0.20}$    \\
118      & 67066    &  214.9830245  &  52.9560011   &  7.48$^{+0.05}_{-0.04}$  &  10.58$^{+0.05}_{-0.05}$  &  2.14$^{+0.05}_{-0.05}$  &  1.52$^{+0.06}_{-0.06}$    \\
119      & 67073    &  214.8001303  &  52.8232104   &  2.80$^{+0.12}_{-0.19}$  &  10.31$^{+0.08}_{-0.11}$  &  1.42$^{+0.22}_{-1.49}$  &  1.79$^{+0.23}_{-0.84}$    \\
120      & 67919    &  214.9440410  &  52.9297441   &  2.47$^{+3.16}_{-0.08}$  &  9.43$^{+0.47}_{-0.20}$  &  0.86$^{+0.74}_{-0.24}$  &  2.69$^{+0.26}_{-0.76}$    \\
121      & 68963    &  214.9315628  &  52.9210090   &  2.49$^{+0.22}_{-0.09}$  &  9.89$^{+0.15}_{-0.13}$  &  1.14$^{+0.20}_{-0.15}$  &  3.71$^{+0.19}_{-0.32}$    \\
122$^\dagger$      & 69075    &  215.0084905  &  52.9779735   &  7.96$^{+0.07}_{-0.05}$  &  9.45$^{+0.06}_{-0.06}$  &  1.53$^{+0.06}_{-0.05}$  &  1.47$^{+0.10}_{-0.08}$    \\
123      & 69084    &  214.9774708  &  52.9534870   &  3.76$^{+0.09}_{-0.70}$  &  10.94$^{+0.14}_{-0.14}$  &  2.43$^{+0.21}_{-0.33}$  &  3.72$^{+0.20}_{-0.21}$    \\
124      & 69557    &  214.9257560  &  52.9185250   &  3.59$^{+0.10}_{-0.15}$  &  10.58$^{+0.08}_{-0.13}$  &  1.98$^{+0.13}_{-0.14}$  &  3.22$^{+0.25}_{-0.16}$    \\
125      & 69697    &  214.8890698  &  52.8926163   &  3.60$^{+0.09}_{-0.07}$  &  10.63$^{+0.09}_{-0.13}$  &  2.12$^{+0.15}_{-0.14}$  &  3.32$^{+0.17}_{-0.17}$    \\
126      & 70195    &  214.8505690  &  52.8660278   &  3.11$^{+0.09}_{-0.09}$  &  10.64$^{+0.07}_{-0.17}$  &  2.01$^{+0.18}_{-0.14}$  &  2.80$^{+0.30}_{-0.21}$    \\
127$^\dagger$      & 71049    &  214.8400344  &  52.8606505   &  4.75$^{+0.91}_{-0.15}$  &  9.74$^{+0.06}_{-0.09}$  &  1.25$^{+0.14}_{-0.12}$  &  1.42$^{+0.17}_{-0.12}$    \\
128      & 71055    &  214.8790993  &  52.8880654   &  3.02$^{+0.02}_{-0.03}$  &  9.85$^{+0.16}_{-0.05}$  &  1.89$^{+0.03}_{-0.07}$  &  2.36$^{+0.06}_{-0.19}$    \\
129      & 71122    &  215.0390567  &  53.0027819   &  3.22$^{+0.15}_{-0.12}$  &  10.45$^{+0.08}_{-0.07}$  &  1.58$^{+0.13}_{-0.58}$  &  2.44$^{+0.24}_{-0.65}$    \\
130      & 72378    &  215.0215373  &  52.9913009   &  2.70$^{+1.71}_{-0.19}$  &  10.95$^{+0.42}_{-0.09}$  &  2.17$^{+0.95}_{-0.17}$  &  3.90$^{+0.07}_{-0.65}$    \\
131      & 73426    &  214.8670444  &  52.8832805   &  3.54$^{+0.11}_{-0.34}$  &  10.36$^{+0.10}_{-0.10}$  &  2.00$^{+0.14}_{-0.21}$  &  1.92$^{+0.15}_{-0.17}$    \\
132$^\dagger$      & 73685    &  214.9233729  &  52.9255931   &  7.27$^{+0.07}_{-0.07}$  &  9.85$^{+0.06}_{-0.05}$  &  1.93$^{+0.04}_{-0.05}$  &  1.96$^{+0.08}_{-0.06}$    \\
133      & 73705    &  214.8013661  &  52.8370353   &  3.64$^{+0.05}_{-0.07}$  &  9.80$^{+0.09}_{-0.06}$  &  1.85$^{+0.04}_{-0.04}$  &  2.30$^{+0.06}_{-0.08}$    \\
134      & 73825    &  215.0045564  &  52.9835262   &  3.64$^{+0.08}_{-0.08}$  &  10.70$^{+0.10}_{-0.15}$  &  2.22$^{+0.14}_{-0.12}$  &  3.48$^{+0.17}_{-0.18}$    \\
135      & 74051    &  214.7856935  &  52.8258160   &  2.07$^{+0.37}_{-0.24}$  &  10.22$^{+0.15}_{-0.17}$  &  1.72$^{+0.23}_{-0.38}$  &  3.39$^{+0.20}_{-0.35}$    \\
136$^\dagger$      & 74228    &  214.9724417  &  52.9621923   &  7.24$^{+0.14}_{-0.11}$  &  9.86$^{+0.11}_{-0.11}$  &  1.64$^{+0.14}_{-0.13}$  &  1.98$^{+0.21}_{-0.20}$    \\
137      & 74393    &  214.8657827  &  52.8834206   &  1.61$^{+1.18}_{-0.21}$  &  9.57$^{+0.30}_{-0.15}$  &  0.57$^{+0.64}_{-0.21}$  &  2.75$^{+0.33}_{-0.50}$    \\
138      & 75238    &  214.7680280  &  52.8163996   &  3.59$^{+0.09}_{-0.06}$  &  10.54$^{+0.10}_{-0.12}$  &  2.03$^{+0.15}_{-0.14}$  &  3.49$^{+0.21}_{-0.18}$    \\

\end{tabular} 
\end{table*}

\begin{table*}
 
\contcaption{}
\centering
\label{tab:continued3}

\begin{tabular}{|c|c|c|c|c|c|c|c|}
\hline
S.No. & ID & RA & Dec & $z_\mathrm{{phot}}$ & $\mathrm{\log\ M_\star/M_\odot}$ & $\mathrm{\log\ SFR/M_\odot yr^{-1}}$ & $\mathrm{A_V\ /mag}$ \\
\hline
\hline

139      & 76999    &  214.7672283  &  52.8177106   &  3.02$^{+0.02}_{-0.02}$  &  10.04$^{+0.06}_{-0.03}$  &  2.10$^{+0.02}_{-0.02}$  &  2.72$^{+0.04}_{-0.07}$    \\
140      & 77220    &  214.8396806  &  52.8717324   &  2.96$^{+0.12}_{-0.13}$  &  9.73$^{+0.06}_{-0.12}$  &  1.03$^{+0.16}_{-0.12}$  &  1.79$^{+0.22}_{-0.17}$    \\
141      & 78330    &  214.7914982  &  52.8380321   &  2.23$^{+0.49}_{-0.14}$  &  10.93$^{+0.13}_{-0.08}$  &  2.02$^{+0.27}_{-0.16}$  &  3.84$^{+0.11}_{-0.33}$    \\
142      & 79082    &  214.9183809  &  52.9378937   &  2.80$^{+0.18}_{-0.18}$  &  9.89$^{+0.08}_{-0.16}$  &  1.20$^{+0.19}_{-0.14}$  &  2.14$^{+0.26}_{-0.19}$    \\
143      & 79446    &  214.8351098  &  52.8951289   &  5.15$^{+0.05}_{-0.05}$  &  9.99$^{+0.09}_{-0.08}$  &  1.78$^{+0.06}_{-0.08}$  &  1.57$^{+0.08}_{-0.11}$    \\
144      & 79727    &  214.9257694  &  52.9544458   &  2.06$^{+1.74}_{-0.26}$  &  9.63$^{+0.41}_{-0.17}$  &  0.78$^{+0.65}_{-0.30}$  &  3.49$^{+0.35}_{-0.79}$    \\
145      & 80544    &  215.0112724  &  53.0135961   &  5.18$^{+0.04}_{-0.04}$  &  10.22$^{+0.05}_{-0.05}$  &  2.26$^{+0.03}_{-0.03}$  &  1.94$^{+0.05}_{-0.04}$    \\
146$^\dagger$      & 80697    &  214.7598250  &  52.8334125   &  3.63$^{+0.62}_{-1.97}$  &  9.46$^{+0.24}_{-0.56}$  &  0.96$^{+0.20}_{-1.00}$  &  2.63$^{+1.14}_{-0.39}$    \\
147      & 81918    &  214.9145423  &  52.9430232   &  4.79$^{+0.05}_{-0.07}$  &  9.12$^{+0.09}_{-0.06}$  &  1.08$^{+0.05}_{-0.09}$  &  1.74$^{+0.11}_{-0.17}$    \\
148      & 82924    &  214.9091113  &  52.9372134   &  3.68$^{+0.54}_{-0.39}$  &  10.10$^{+0.18}_{-0.16}$  &  1.48$^{+0.27}_{-0.23}$  &  3.60$^{+0.24}_{-0.26}$    \\
149      & 83296    &  214.9040279  &  52.9327056   &  2.10$^{+0.44}_{-0.15}$  &  10.53$^{+0.10}_{-0.10}$  &  1.61$^{+0.20}_{-0.18}$  &  3.48$^{+0.25}_{-0.37}$    \\
150$^\dagger$      & 83338    &  214.9508401  &  52.9668645   &  3.83$^{+0.34}_{-0.23}$  &  9.64$^{+0.11}_{-0.21}$  &  1.11$^{+0.14}_{-0.15}$  &  2.36$^{+0.27}_{-0.22}$    \\
151      & 83822    &  214.7665808  &  52.8315226   &  4.30$^{+0.17}_{-0.13}$  &  9.64$^{+0.07}_{-0.10}$  &  1.14$^{+0.13}_{-0.11}$  &  1.82$^{+0.18}_{-0.15}$    \\
152      & 83936    &  214.9491882  &  52.9641429   &  6.20$^{+0.20}_{-0.29}$  &  9.46$^{+0.09}_{-0.10}$  &  1.52$^{+0.08}_{-0.09}$  &  2.09$^{+0.13}_{-0.13}$    \\
153$^\dagger$      & 84323    &  214.9257531  &  52.9456643   &  7.27$^{+0.11}_{-0.11}$  &  9.83$^{+0.12}_{-0.12}$  &  1.84$^{+0.11}_{-0.11}$  &  2.48$^{+0.19}_{-0.18}$    \\
154      & 84655    &  214.8383963  &  52.8851887   &  6.07$^{+0.27}_{-0.23}$  &  9.60$^{+0.07}_{-0.07}$  &  1.67$^{+0.07}_{-0.06}$  &  1.82$^{+0.06}_{-0.07}$    \\
155      & 85172    &  214.8109343  &  52.8589270   &  3.01$^{+0.31}_{-0.18}$  &  9.62$^{+0.09}_{-0.17}$  &  0.90$^{+0.17}_{-0.17}$  &  1.90$^{+0.30}_{-0.23}$    \\
156      & 85249    &  214.9887041  &  52.9886234   &  1.71$^{+1.90}_{-0.11}$  &  9.82$^{+0.43}_{-0.14}$  &  0.94$^{+0.93}_{-0.22}$  &  2.70$^{+0.28}_{-0.73}$    \\
157      & 85675    &  214.9469656  &  52.9602699   &  1.95$^{+0.18}_{-0.47}$  &  10.01$^{+0.14}_{-0.16}$  &  1.24$^{+0.27}_{-0.34}$  &  2.66$^{+0.29}_{-0.24}$    \\
158      & 87151    &  214.8485472  &  52.8847626   &  1.84$^{+1.96}_{-0.19}$  &  9.08$^{+0.53}_{-0.15}$  &  0.18$^{+0.91}_{-0.21}$  &  3.53$^{+0.33}_{-1.03}$    \\
159      & 87239    &  214.8111763  &  52.8586480   &  2.18$^{+1.49}_{-0.56}$  &  10.33$^{+0.14}_{-0.30}$  &  1.78$^{+0.69}_{-0.59}$  &  3.34$^{+0.45}_{-0.52}$    \\
160      & 87370    &  214.7792320  &  52.8369189   &  5.93$^{+1.89}_{-2.17}$  &  9.22$^{+0.19}_{-0.46}$  &  0.97$^{+0.21}_{-0.61}$  &  1.74$^{+0.49}_{-0.36}$    \\
161      & 87446    &  214.8751893  &  52.9134883   &  2.96$^{+0.08}_{-0.10}$  &  10.01$^{+0.09}_{-0.15}$  &  1.46$^{+0.15}_{-0.15}$  &  2.07$^{+0.20}_{-0.19}$    \\
162      & 88423    &  214.8482919  &  52.8847861   &  3.94$^{+0.19}_{-2.14}$  &  9.58$^{+0.15}_{-0.46}$  &  1.25$^{+0.17}_{-0.93}$  &  2.00$^{+0.78}_{-0.23}$    \\
163      & 88428    &  214.7634081  &  52.8477946   &  1.77$^{+0.11}_{-0.11}$  &  10.14$^{+0.09}_{-0.11}$  &  1.10$^{+0.16}_{-0.27}$  &  3.74$^{+0.17}_{-0.22}$    \\
164      & 90408    &  214.8293099  &  52.8939285   &  2.53$^{+0.16}_{-0.32}$  &  11.22$^{+0.06}_{-0.07}$  &  1.84$^{+0.53}_{-1.21}$  &  3.59$^{+0.27}_{-0.53}$    \\
165      & 90510    &  214.7437385  &  52.8368225   &  3.78$^{+0.04}_{-1.25}$  &  10.58$^{+0.09}_{-0.51}$  &  2.62$^{+0.04}_{-0.49}$  &  2.69$^{+0.28}_{-0.11}$    \\
166      & 91433    &  214.8520783  &  52.9097650   &  2.24$^{+0.08}_{-0.09}$  &  10.44$^{+0.07}_{-0.10}$  &  1.60$^{+0.15}_{-0.11}$  &  3.01$^{+0.27}_{-0.23}$    \\
167      & 92377    &  214.8918896  &  52.9338667   &  3.96$^{+1.78}_{-0.19}$  &  10.36$^{+0.27}_{-0.23}$  &  1.93$^{+0.39}_{-0.23}$  &  3.10$^{+0.35}_{-0.46}$    \\
168      & 93595    &  214.9578855  &  52.9802999   &  3.03$^{+0.02}_{-0.02}$  &  10.02$^{+0.07}_{-0.03}$  &  2.07$^{+0.02}_{-0.02}$  &  2.48$^{+0.04}_{-0.06}$    \\

\end{tabular} 
\end{table*}


\bsp	
\label{lastpage}
\end{document}